\documentclass[aps,twocolumn]{revtex4}
\makeatletter

\newcommand{\Rmnum}[1]{\expandafter\@slowromancap\romannumeral #1@}
\makeatother
\usepackage{graphicx}
\usepackage{dcolumn}
\usepackage{bm}
\usepackage{CJK}
\usepackage{color}
\usepackage{amsfonts}
\usepackage{psfrag}
\usepackage{wrapfig}
\usepackage{subfigure}
\usepackage{makeidx}
\usepackage{multirow}
\usepackage{epsf}
\usepackage[colorlinks,linkcolor=blue,anchorcolor=red,citecolor=blue,urlcolor=blue]{hyperref}
\usepackage{amsmath}
\usepackage{cases}

\begin{document}

\title{Growth rate of modulation instability driven by superregular breathers}
\author{Chong Liu$^{1,2,3}$}\email{chongliu@nwu.edu.cn}
\author{Zhan-Ying Yang$^{1,2}$}\email{zyyang@nwu.edu.cn}
\author{Wen-Li Yang$^{1,2,4}$}\email{wlyang@nwu.edu.cn}
\address{$^1$School of Physics, Northwest University, Xi'an 710069, China}
\address{$^2$Shaanxi Key Laboratory for Theoretical Physics Frontiers, Xi'an 710069, China}
\address{$^3$Optical Sciences Group, Research School of Physics and Engineering, The Australian National University, Canberra, ACT 2600, Australia}
\address{$^4$Institute of Modern Physics, Northwest University, Xi'an 710069, China}
\begin{abstract}
We report an exact link between Zakharov-Gelash super-regular (SR) breathers (formed by a pair of quasi-Akhmediev breathers) with interesting  different nonlinear propagation characteristics and modulation instability (MI).
This shows that the absolute difference of group velocities of SR breathers coincides exactly with the linear MI growth rate.
This link holds for a series of nonlinear Schr\"{o}dinger equations with infinite-order terms.
For the particular case of SR breathers with opposite group velocities, the growth rate of SR breathers is consistent with that of each quasi-Akhmediev breather along the propagation direction.
Numerical simulations reveal the robustness of different SR breathers generated from various non-ideal single and multiple initial excitations.
Our results provide insight into the MI nature described by SR breathers and could be helpful for controllable SR breather excitations
in related nonlinear systems.
\end{abstract}

\maketitle
\textbf{Modulation instability (MI), i.e., the instability of a constant background
with respect to to small periodic or irregular perturbations, is the central process in many different
physical settings ranging from hydromechanics, nonlinear optics, plasma, and Bose-Einstein condensates.
Although the property of initial linear MI stage (i.e., the MI criterion and growth rate) can be studied well by the linear stability analysis, rigorous analytic descriptions of the complete MI process involving both linear and nonlinear stages are
long-standing challenges and thus are far less common.
So far, only a few analytic descriptions of MI have been obtained including the well-known Akhmediev breather and Peregrine breather, as well as the newly proposed super-regular (SR) breather. One crucial step in the analytic descriptions of MI is to establish the exact link between them and the linear MI to show the nature of MI quantitatively.
Here we report an exact link between SR breathers and linear MI
by showing that the absolute difference of group velocities of SR breathers coincides exactly with the linear MI growth rate.
This link holds for a greatly broad domain described by the nonlinear Schr\"{o}dinger equations up to infinite order. Numerical simulations are carried out to confirm not only the exact result but also the robustness of different nontrivial SR breathers generated from various non-ideal single and multiple initial excitations.
}
\section{Introduction}
Modulation instability (MI) \cite{MI0} is a central process of nonlinear physics; it  has attracted particular
attention in both theory and experiment, since it is regarded as the origin of solitons \cite{MIs}, supercontinuum generation \cite{MI2},
and rogue wave events \cite{MIr0,MI1,MIr1,MIN1,MIN2,MIr2,MIr3}. In general, a complete MI process consists of an initial linear stage
(where weak (small-amplitude) perturbations suffer exponential growth on a constant background field) and a subsequent nonlinear evolution stage with rich dynamics.
In general, the property of initial linear MI stage (i.e., the instability criterion and growth rate)
can be successfully studied by linear stability analysis. Nonetheless, rigorous analytic descriptions of the complete MI process are
long-standing challenges and thus are far less common \cite{MInews}.

The first intriguing analytic MI prototype is known as the Akhmediev breather (AB) \cite{AB1,ABb}. The latter is an exact
solution of the standard nonlinear Schr\"{o}dinger equation (NLSE), which describes MI evolution from
weak periodic (infinite-width) perturbations and exhibits one growth-return cycle. This has been interpreted strictly
by the exact link between an AB solution and  linear MI (see the geometric interpretation in Ref. \cite{ABb}).
Thus,   AB is referred as an elementary MI and provides \textit{a new analysis of an old instability} \cite{MInews}.
Indeed, this has led to successful studies of the MI spectral characteristics \cite{AB2}, experiments exciting the Peregrine
rogue wave (RW) \cite{PRW}  and the revelation of higher-order MI \cite{MIH1}.

Nevertheless, exact descriptions beyond the AB are both relevant and necessary
in order to model more general and important MI scenarios from \textit{localized weak perturbations}.
The exact solution of Peregrine RW \cite{PRW1} (the periodic limitation of ABs) is thus considered as the simplest rigorous description
of a MI process from localized single-peak weak perturbations. Surprisingly, despite the worldwide success, in particular, stimulating
new theoretical and experimental studies on the rich family of Peregrine-type solutions \cite{MIr0,MIr1,MI1,MIN1,MIN2,MIr2,MIr3}, the MI nature
of the Peregrine RW is
strictly revealed only recently by establishing a crucial exact link between the Peregrine RW solution and MI \cite{MIr2}.
This shows exactly that the Peregrine RW describes a particular MI scenario corresponding to the zero-frequency sub-region,
which substantiates the previously inconclusive qualitative argument.

Another remarkable MI evolution from \textit{localized multi-peak weak perturbations} is the so-called Zakharov-Gelash super-regular (SR) breather \cite{SR1,SR2}.
In contract to the AB and Peregrine RW, this particular MI scenario exhibits complex nonlinear evolution of multiple quasi-ABs,
and is associated with higher-order MI \cite{MIH1}.
Recently, significant progress has been made on SR breathers, including the dynamical observation of them in both optics
and hydrodynamics \cite{SR3}, their utility in generating rogue wave events \cite{SR4}, and the excitation of noise-driven SR breathers [see Chap. 7 in Ref.\cite{SR5}].
However, an exact link between SR breathers and MI has not been established fully.
The difficulty could stem from that the SR breather solution is a complex higher-order solution, a valid analysis in complex nonlinear systems is scarce.
In this paper,
we reveal a hidden exact link between SR breathers and MI \textit{with universality}
by showing that the absolute difference of group velocities of SR breathers coincides with the linear MI growth
rate.
To this end, we shall not consider a special NLSE case; instead we go beyond the standard NLSE by considering a generalized NLSE{---}the infinite NLSE hierarchy \cite{HA1}
\begin{eqnarray}
iu_\xi+\sum_{n=1}^{\infty}\left[\alpha_{2n} K_{2n}(u)-i\alpha_{2n+1} K_{2n+1}(u)\right]=0,\label{eqin}
\end{eqnarray}
where $u(\tau,\xi)$ is the complex field, $\xi$ and $\tau$ are the longitudinal and transverse variables, respectively. Each coefficient $\alpha_n, n=2,3,4,5, ... ,\infty$, is an arbitrary real number which is responsible for the different-order dispersion and nonlinear terms.
Eq. (\ref{eqin}) is an important integrable extension of NLSEs up to infinite order \cite{HA1,HA2,HA3}.
Specifically, $K_2(u)$ is the second-order NLSE term \cite{NLSE1}: $K_2(u)=u_{\tau\tau}+2|u|^2u$;
$K_3(u)$ is the third-order term with third-order dispersion \cite{NLSE2}: $K_3(u)=u_{\tau\tau\tau}+6|u|^2u_\tau$; $K_4(u)$ is the
fourth-order term with fourth-order dispersion \cite{NLSE3}:
$K_4(u)=u_{\tau\tau\tau\tau}+6u_\tau^2u^\ast+4|u_\tau|^2u+8|u|^2u_{\tau\tau}+2u^2u_{\tau\tau}^\ast+6|u|^4u$;
$K_5(u)$ is the quintic term with fifth-order dispersion \cite{NLSE4}:
$K_5(u)=u_{\tau\tau\tau\tau\tau}+10|u|^2u_{\tau\tau\tau}+10(|u_\tau|^2u)_\tau+20u^\ast u_{\tau}u_{\tau\tau}+30|u|^4u_{\tau}$.
The next higher-order terms are given by the formula in \cite{HA1}. To determine the common and different characteristics of
odd and even terms, individually or in combination with
the basic NLSE term, we will first consider the first five
terms and then we
extend the result into the infinite NLSE (\ref{eqin}).

Physically, the infinite NLSE (\ref{eqin}) is usually considered to be improved models
for a more accurate description of nonlinear wave propagation in the ocean and in optical fibers [see Chap. 10 in \cite{SR5}].
Indeed, it has been referred as a special integrable case of a more general governing equation for pulse propagation in an optical fiber \cite{Heo}.
On the other hand, recent studies demonstrate that higher-order NLSEs are of practical significance in the dynamical
description of water waves \cite{Hew1}; in particular, they describe the experimental results collected in
water tanks with higher accuracy than the standard NLSE \cite{Hew}.
Thus, the study of nonlinear waves with each higher-order term or any combination of the terms will be a significant progress
in both theory and experiment \cite{HEq0,HEq,HEq1,HEq2,He,H1,H2,H3,H4,H5,Hyan,H6,H7,H8,H9,H10}.

Theoretically, the fundamental (first-order) solution and second-order rational solution on a plane-wave background in the infinite NLSE (\ref{eqin}) has been obtained \cite{HA1,HA2,HA3}. Considering some special finite terms (up to fifth order), various nonlinear structures, on both zero and
nonzero backgrounds, have previously been presented \cite{HEq0,HEq,HEq1,HEq2,H1,H2,H3,Hyan,H4,H5,H6,H7,H8,H9,H10}.
A remarkable feature is that nonlinear waves on a plane-wave background can exhibit structural
diversity induced by higher-order effects \cite{H1,H2,H3,H4,H5,H6,Hyan,H7,H8,H9}.
Consequently some new types of nonlinear modes have been revealed.
However, as a special type of higher-order breather of physical importance, SR breathers in the infinite NLSE (\ref{eqin}) remain unexplored so far;
in particular, the exact link between SR breathers and MI has not been investigated completely.
Indeed, up to now, only the existence of SR breathers was demonstrated in the particular case by considering the single third or fourth order term \cite{HSR1,HSR2}.
It is expected that the study of the generalized physical model (\ref{eqin}) will yield results with universality.
In the following, we confine our attention to the general property of SR breathers with all non-vanishing higher-order terms.
The remaining special cases, with each higher-order term, can be studied readily from the general results.

\section{SR breather property}
We first consider SR breather property in the first five equations, i.e., the higher-order NLSE up to fifth order.
The analytic SR breather solution is constructed by the Darboux transformation, but the spectral
parameter $\lambda$ is parametrized by the Jukowsky transform \cite{SR1,SR2},
\begin{equation}
\lambda=i\frac{a}{2}\left(\Delta+\frac{1}{\Delta}\right)-\frac{q}{2},~\Delta=Re^{i\phi}.\label{eqla}
\end{equation}
Here $R$, $\phi$ are the radius and angle of the polar coordinates in the region $R\geq1$, $\phi\in(-\pi/2,\pi/2)$.
In contrast to the conventional $\lambda$, one can establish a concise phase diagram of many different nonlinear modes
in the polar coordinate ($Re[\Delta]-Im[\Delta]$) plane \cite{SR2,HSR2}. For the higher-order solution, Eq. (\ref{eqla}) provides
a new type of breather collisions with a localized weak structure at a certain $\xi$ when $R\rightarrow1$, which is the so-called SR breather.
The real parameters $a$ and $q$ denote the amplitude and frequency of the initial plane-wave background:
\begin{equation}
u_0=ae^{i\theta},~~\theta=q \tau+\omega \xi,\label{eqb}
\end{equation}
where $\omega=\alpha_2\left(2a^2-q^2\right)+\alpha_3\left(6a^2 q-q^3\right)+\alpha_4\left(6a^4-12a^2 q^2+q^4\right)+\alpha_5\left(30 a^4 q-20 a^2 q^3+q^5\right)$.
If $q=0$, $\lambda$ reduces to the simplest case in \cite{SR1} and $u_0$ degenerates to the case in \cite{HEq1}.
In fiber optics, the parameter $q$ represents the frequency of pump wave \cite{Heo}. For the simplest NLSE, changing the value of $q$ has no essential influence on breather dynamics, since it can be eliminated from the solution by the Galilean transformation
(see, e.g., \cite{GT}). However, if one considers  nonlinear systems beyond the description of the simplest NLSE (where the Galilean transformation is broken), the pump wave frequency provides an additional degree of freedom for generating nontrivial breather dynamics (see below and also \cite{H1,q,v5}).

Collecting Eqs. (\ref{eqla}) and (\ref{eqb}) into the associated Lax pair in Ref. \cite{HEq}, we can obtain a
second-order solution with the parameters of $\lambda_j$: $\phi_1=-\phi_2=\phi$, $R_1=R_2=R=1+\varepsilon$,
where $\varepsilon$ is a small value ($\varepsilon\ll1$) [see Appendix A].
The solution depends on $a$, $q$, $\alpha_2$, $\alpha_3$, $\alpha_4$, $\alpha_5$, $\varepsilon$, $\phi$, and the additional
phase parameters $\theta_j$, $\mu_j$ ($j=1,2$).
The latter plays a key role in the formation of SR breathers. Here we choose $\theta_{1}+\theta_{2}=\pi$, $\mu_{1,2}=0$ \cite{SR1,SR2},
implying that the small perturbation $\delta u$ (see Fig. \ref{figso1}) emerges at $\xi=0$ [i.e., $u(0,\tau)=(a+\delta u)e^{i\theta}$].
Interestingly, the small perturbation $\delta u$ can be obtained explicitly by the procedure described in Refs. \cite{SR1,SR2}; it  reads
\begin{equation}
\delta u\approx-i\left(\frac{4a\varepsilon\cosh{(i\phi)}}{\cosh{(2a\varepsilon \tau \cos{\phi})}}\right)\cos{\left(2a\tau\sin{\phi}\right)}.
\label{eqp}
\end{equation}
Remarkably, $\delta u$ is a purely imaginary small-amplitude perturbation which consists of a localized
function and a periodic modulation function with frequency $2a\sin{\phi}$ [or more precisely, $a(R+1/R)\sin{\phi}$].
The width of $\delta u$ can be
estimated as  $1/(2a\varepsilon\cos{\phi})$. Thus $\delta u$ is a localized form with broad edges since $\varepsilon\ll1$.
Interestingly, Eq. (\ref{eqp}) only depends on the parameters $a$, $\varepsilon$, $\phi$. It has no connection with the
structure parameters ($\alpha_2$, $\alpha_3$, $\alpha_4$, $\alpha_5$) of Eq. (\ref{eqin}) as well as the background frequency $q$ in Eq. (\ref{eqb}). This indicates that $\delta u$
remains invariant in NLSEs with different-order terms. Indeed, Eq. (\ref{eqp}) is valid for the whole infinite NLSEs (\ref{eqin}).
Thus different nonlinear stages of SR waves
can  evolve  from an identical initial perturbation $\delta u$ with fixed $a$, $R$, $\phi$. In the following, our interest is focused on the properties of different SR breathers from \textit{an identical initial state}.

\begin{figure}[htb]
\centering
\includegraphics[height=32mm,width=84mm]{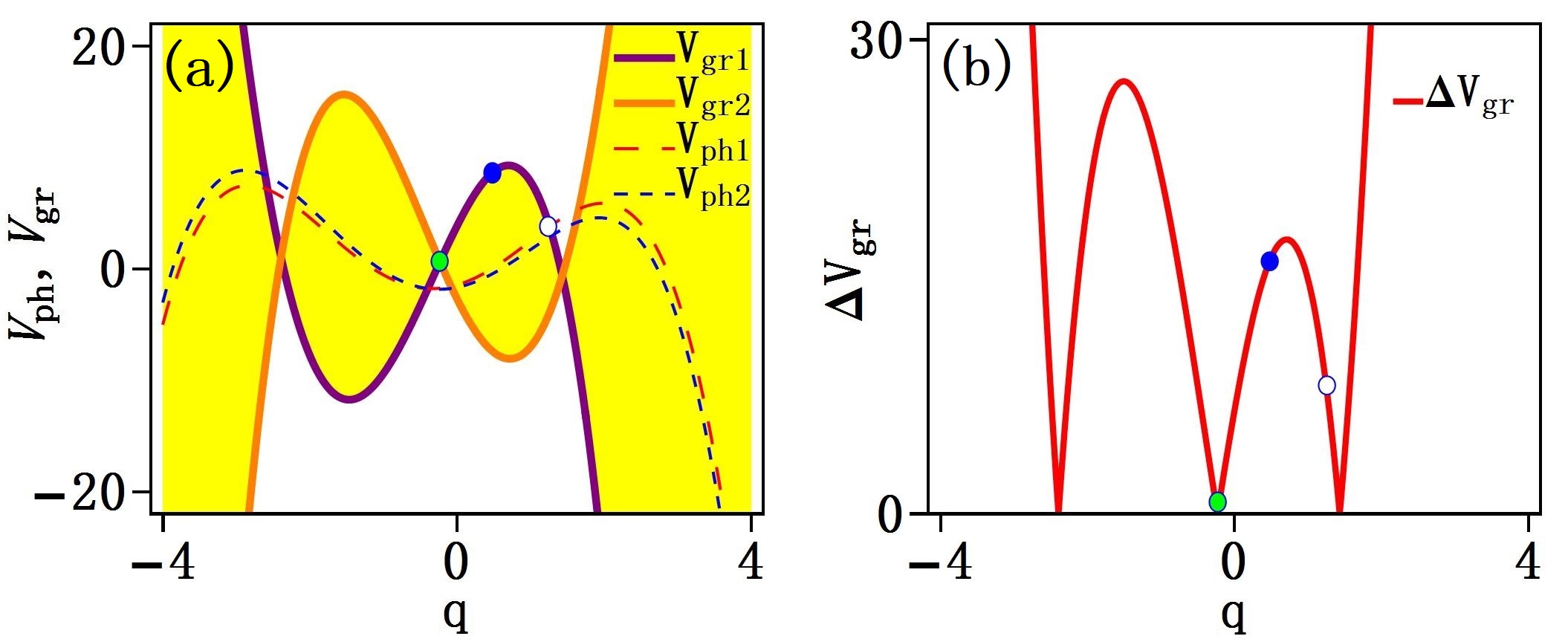}
\caption{(a) Evolution of $V_{grj}$, $V_{phj}$ of SR breathers with $q$ [the exact expression in Appendix A]. The yellow region represents the group velocity difference: $V_{gr1}-V_{gr2}$. (b) Evolution of the absolute difference of group velocities: $\Delta V_{gr}=|V_{gr1}-V_{gr2}|$. The blue, white,
green dots correspond to $q=0.5,q_t,q_s$, which are extracted from conditions: $\Delta V_{gr}\neq0$, $V_{grj}\neq V_{phj}$; $\Delta V_{gr}\neq0$, $V_{grj}=V_{phj}$,
$V_{gr3-j}\neq V_{ph3-j}$; $\Delta V_{gr}=0$.
They describe different SR states as $\Delta V_{gr}\rightarrow0$ in Figs. \ref{figso1} (a), (b), (c), respectively.
The setup is: $a=1, \alpha_2=0, \alpha_3=0.2, \alpha_4=0.1, \alpha_5=0.05$, $\phi=\pi/4, R=1.2, \theta_{1,2}=\pi/2, \mu_{1,2}=0$.
}\label{figso}
\end{figure}

\begin{figure*}[htb]
\centering
\includegraphics[height=60mm,width=128mm]{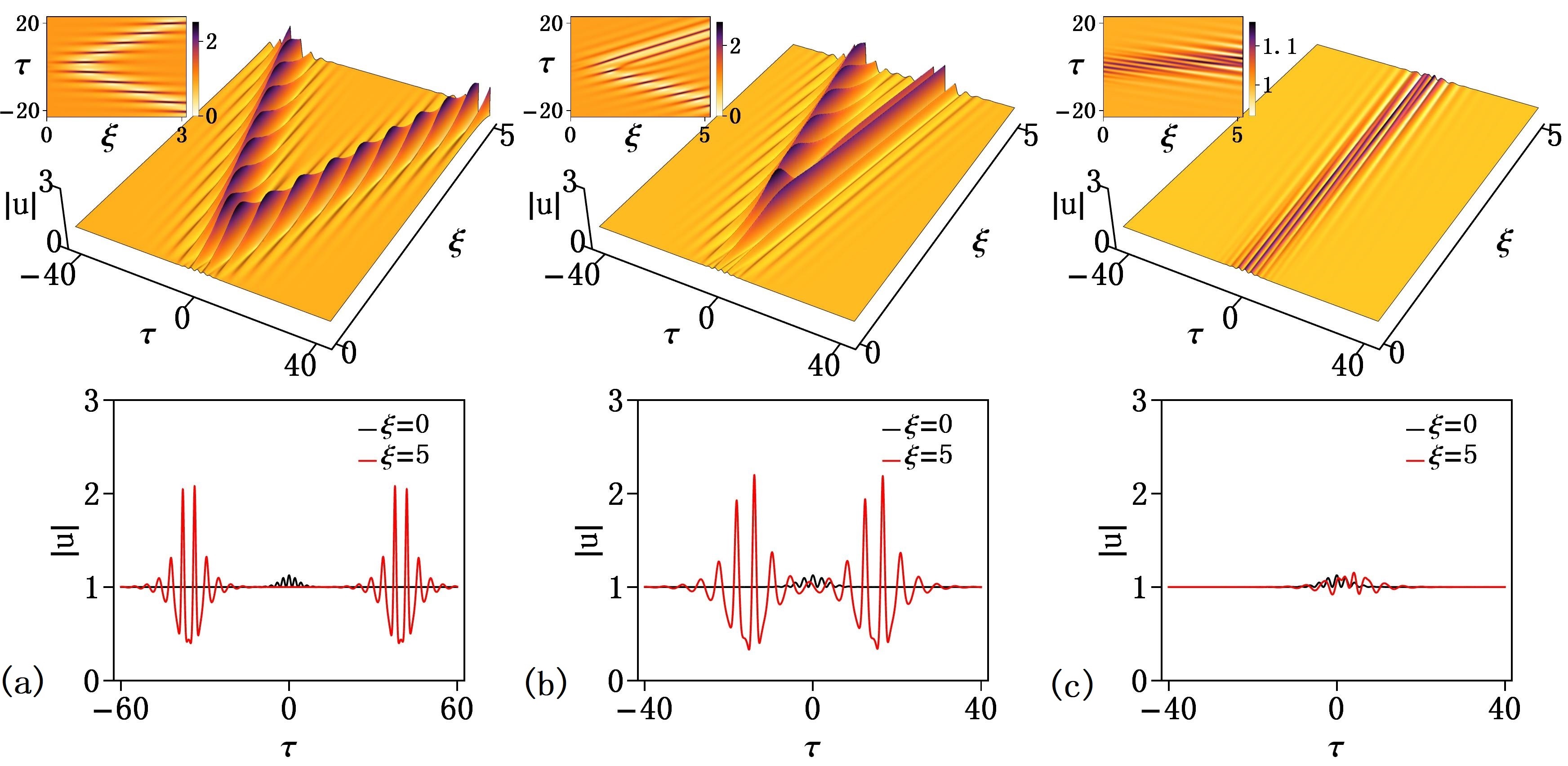}
\caption{Amplitude distributions $|u|$ of different SR states excited from an identical initial state as $\Delta V_{gr}\rightarrow0$.
Here (a) shows the standard SR breathers with $\Delta V_{gr}\neq0$, $V_{grj}\neq V_{phj}$, thus we choose $q=0.5$
[the blue dot in Fig. \ref{figso}]; (b) displays the half-transition SR state with $\Delta V_{gr}\neq0$, $V_{grj}=V_{phj}$,
$V_{gr3-j}\neq V_{ph3-j}$ ($q=q_t$) [the white dot in Fig. \ref{figso}]; (c) presents the SR bound state with $\Delta V_{gr}=0$
($q=q_s$) [the green dot in Fig. \ref{figso}]. The initial perturbation at $\xi=0$ is described by Eq. (\ref{eqp}).
The setup is same as in Fig. \ref{figso}.
}\label{figso1}
\end{figure*}

However, a property analysis of nonlinear evolution of SR breathers in Eq. (\ref{eqin}) is not an easy task, since
multiple parameters are involved in the solution.
Nevertheless, this obstacle can be overcomed when one focuses on the main factors that determine the physical
properties of SR breathers, i.e., the group and phase velocities $V_{grj}$, $V_{phj}$.
These explicit expressions come from the exact SR breather solution (see Appendix A), where
$V_{grj}$ is extracted from the hyperbolic function $\cosh\Theta_j$, while $V_{phj}$ is extracted from the trigonometric function $\cos\Phi_j$.
If the higher-order terms are absent, $\alpha_{n>2}=0$, $V_{grj}$, $V_{phj}$ reduce to their values for  the simplest NLSE \cite{SR1,SR2,SR3,SR4}.

Figure \ref{figso} shows the evolution of $V_{grj}$, $V_{phj}$ with $q$.
The complexity of $V_{grj}$ and $V_{phj}$ stems from the higher-order effects.
This will result in nontrivial SR modes which are absent in the standard NLSE [see Figs. \ref{figso1}(b) and (c)].
Importantly, the existence conditions of different SR states can be obtained exactly. Especially,
we define the absolute difference of group velocities in order to understand better the properties of SR waves:
\begin{equation}
\Delta V_{gr}=\left|V_{gr1}-V_{gr2}\right|.\label{eqv}
\end{equation}
One can then confirm readily that when the initial state is chosen (i.e., $R$, $\phi$ are fixed), the standard NLSE allows a constant $\Delta V_{gr}$ as $q$ varies.
This indicates that SR breathers in this case are consistent with each other. The underlying physical
interpretation is that the linear growth rate remains fixed as $q$ varies. It was shown that SR breather
corresponds to the peak MI growth rate when $\phi=\pi/4$ \cite{SR4}, which coincides with that of ABs \cite{AB1}.
However, $\Delta V_{gr}$ in Eq. (\ref{eqin}) with first five terms is nontrivial which can result in many interesting properties of SR breathers.
Here, we present the explicit general expression of $\Delta V_{gr}$:
\begin{eqnarray}
\Delta V_{gr}&=&4Aa|\{\alpha_2 +3q\alpha_3+2(a^2 B\cos2\phi+2a^2-3q^2)\alpha_4\nonumber\\
&&+10q(a^2 B\cos2\phi+2 a^2-q^2)\alpha_5\}\sin\phi|,\label{eqv11}
\end{eqnarray}
where 
$A=(R^4+1)/(R^3-R)$, $B=(R^8+1)/(R^6+R^2)$. Considering the SR wave condition: $R=1+\varepsilon$, $\varepsilon\ll1$, one obtains $A=2/(R-1/R)$, $B=1$. Thus Eq. (\ref{eqv11}) is rewritten as
\begin{eqnarray}
\Delta V_{gr}&=&4Aa|\{\alpha_2 +3q\alpha_3+2(3a^2-3q^2-2a^2 \sin ^2\phi)\alpha_4\nonumber\\
&&+10q(4a^2- q^2-2a^2 \sin ^2\phi)\alpha_5\}\sin\phi|,\label{eqv1}
\end{eqnarray}
Note that Eq. (\ref{eqv1}) is valid for the whole range of $\phi$.
As an example, we depict the evolution of $\Delta V_{gr}$ with $\phi=\pi/4$ in Fig. \ref{figso}(b).
Via an analysis of $V_{grj}$, $V_{phj}$, and $\Delta V_{gr}$, three typical different types of SR states from the identical initial state (\ref{eqp}) with fixed $R$, $\phi$ are shown as $\Delta V_{gr}\rightarrow0$ in Fig. \ref{figso1}.

In Fig. \ref{figso1}(a) we first show a standard SR breather when $\Delta V_{gr}\neq0$, $V_{grj}\neq V_{phj}$
(the blue dot in Fig. \ref{figso}).
As shown, a small localized perturbation at $\xi=0$ is amplified rapidly and ultimately becomes a pair of breathers
propagating  along different directions.
We remark that the solution in this case yields a trivial generalization to SR breathers with higher-order effects.

As $\Delta V_{gr}$ decreases, we can observe an interesting half-transition SR state when $\Delta V_{gr}\neq0$,
$V_{grj}=V_{phj}$, $V_{gr3-j}\neq V_{ph3-j}$, which corresponds to the white dot in Fig. \ref{figso}.
As shown in Fig. \ref{figso1}(b), a breather is converted to a non-breathing wave when $V_{grj}=V_{phj}$, while
the other retains its nature as a breather when $V_{gr3-j}\neq V_{ph3-j}$.
In contrast to the standard SR wave in Fig. \ref{figso1}(a), the small-amplitude perturbation is amplified rather
slowly and subsequently becomes a mix of breathing and non-breathing waves.

Once $\Delta V_{gr}=0$, a novel SR bound state is observed in Fig. \ref{figso1}(c) (the blue dot in Fig. \ref{figso}).
It seems that, in this particular case, the small-amplitude perturbation propagates along $\xi$ with small oscillations,
but the amplification of the perturbation is suppressed completely. Physically, this SR state describes a non-amplifying
nonlinear wave dynamics which corresponds to a vanishing growth rate, $G_{sr}=0$ [see the next Section]. This indicates that
for a specially-designed optical system described by the higher-order NLSEs with fixed structure parameters, one can estimate the pump wave frequency $q=q_s$ to generate the SR bound state by $\Delta V_{gr}=0$, or $G_{sr}=0$.

Remarkably, as shown in Fig. \ref{figso1}, the amplification rate of the identical perturbation $\delta u$
decreases gradually as $\Delta V_{gr}\rightarrow0$. This may correspond to the attenuation of the MI growth rate of the initial
perturbation $\delta u$. In the following, we will give our physical explanation by establishing the exact link between SR breathers
and linear MI growth rate.

On the other hand, we remark that the half-transition and bound-state SR waves appear as a result of the higher-order effects which are absent in the standard NLSE.
Here one can readily check that these nontrivial SR states hold for the cases with each single higher-order term. However, particular attention should be paid on the fourth-order term, in which the symmetry breaking of group velocities occurs when $\alpha_4\neq0$. This leads to the bound-state SR waves with different periodic evolutions (not shown). When $\alpha_4\rightarrow0$, the bound-state SR wave is close to the full-suppression state reported in \cite{HSR2}.
\section{exact link}
MI criterion and growth rate of small-amplitude perturbations on a plane-wave background can be studied
precisely by the linear stability analysis.
A perturbed background $u_p$ is obtained by adding small-amplitude perturbed Fourier modes $p$ on
the background $u_0$, i.e., $u_{p}=[a+p]e^{i\theta}$, where
$p=f_+e^{i(Q \tau+\omega \xi)}+f_{-}^{*}e^{-i(Q \tau+\omega^* \xi)}$ with small amplitudes $f_+$, $f_{-}^{*}$, perturbed
frequency $Q$, and wavenumber $\omega$.
Followed by the standard linearization process, a substitution of $u_p$ into Eq. (\ref{eqin}) with first five terms yields the dispersion relation
 between $\omega$ and $Q$.
The imaginary part of $\omega$ leads to MI:
\begin{eqnarray}
\textrm{Im}\{\omega\}&=&\pm\frac{1}{2}Q\{2\alpha_2 +6 \alpha_3  q -2\alpha_4(Q^2-6 a^2+6 q^2)\nonumber\\
&&-10\alpha_5 [q (Q^2-6 a^2)+2q^3]\}\sqrt{4a^2-Q^2},\label{eqmi}
\end{eqnarray}
where the perturbed frequency satisfies $|Q|<2a$. Interestingly,
the initial state of SR breathers $\delta u$, Eq. (\ref{eqp}), possesses the perturbed frequency $Q_{sr}=2a\sin{\phi}$,
which falls within the MI region $|Q|<2a$.
Thus $\delta u$ is a valid initial state in the MI region which is confirmed exactly by the linear stability analysis.

Let us further consider the linear MI growth rate of the initial state of SR breathers $\delta u$.
Here the growth rate is defined by $G=|\textrm{Im}\{\omega\}|$, which represents the growth rate of amplitude
$|u_p(\tau,\xi)|$. Substituting the initial perturbed frequency $Q_{sr}=2a\sin{\phi}$ in Eq. (\ref{eqp}) into Eq. (\ref{eqmi}),
one can readily obtain the linear MI growth rate for the initial state of SR breathers as follows:
\begin{eqnarray}
G_{sr}&=&2a^2|\{\alpha_2 +3\alpha_3q+2(3a^2-3q^2-2a^2 \sin ^2\phi)\alpha_4\nonumber\\
&&+10q(4a^2- q^2-2a^2 \sin ^2\phi)\alpha_5\}\sin2\phi|.\label{eqg}
\end{eqnarray}
Remarkably, a simple comparison between $G_{sr}$ [Eq. (\ref{eqg})] and $\Delta V_{gr}$ [Eq. (\ref{eqv1})] shows an excellent consistency between them.
Namely, the higher that $G_{sr}$, the larger that $\Delta V_{gr}$ becomes. If $\Delta V_{gr}=0$, the bound state of SR breathers is obtained with $G_{sr}=0$. Note also that due to the higher-order terms ($\alpha_{j>2}$), the peak growth rate and the corresponding dominant frequency will change.
As a result, a hidden exact link between the characteristic quantity of SR breathers and the linear MI is revealed for the
first time; this can be written explicitly as: 
\begin{equation}
G_{sr}=\Delta V_{gr}\cdot\eta_r,\label{eqel}
\end{equation}
where $\eta_r=\frac{a}{2}\left(R-1/R\right)\cos{\phi}$.
This result is important since i) it is an \textit{exact} relation to show the MI property of SR breathers by comparison with
the linear MI; ii) it is a \textit{general} link that holds for the different-order NLSEs.

Let us take a closer look at the exact link (\ref{eqel}) obtained above.
Clearly, when the higher-order terms are absent, i.e., $\alpha_{n>2}=0$, this link (\ref{eqel}) covers that of the simplest NLSE with $q=0$ obtained via the direct derivation of the initial state (\ref{eqp}) [see Eq. (169) in Ref. \cite{SR2}]. This coincidence comes from the fact that the initial state (\ref{eqp}) merely allows the pair of quasi-Akhmediev breathers with symmetric group velocities, i.e., $V_{gr1}=-V_{gr2}$ (thus $\Delta V_{gr}= 2|V_{gr1}|$), in the simplest NLSE with $q=0$. Thus, in this particular case, one can readily obtain $G_{sr}=2\eta_r |V_{gr1}|=\eta_r\Delta V_{gr}=2a^2\alpha_2\sin2\phi$. However, once this symmetry of group velocities is broken (i.e, $V_{gr1}\neq-V_{gr2}$), the direct method in Ref. \cite{SR2}, which is used to obtain the exact link, becomes invalid, even for the simplest NLSE with $q\neq0$ [where the group velocities are unequal, $V_{grj}=2\alpha_2(q\pm2aA\sin\phi)$]. In fact,  unequal group velocities of SR breathers occurs more frequently when one considers   nonlinear systems beyond the description of the simplest NLSE. Moreover, the corresponding dynamics described by these systems
can be highly complicated.
The challenging problem{---}how to establish a general exact link between SR breathers and MI{---}is therefore what we address in this work. Remarkably, the physics of the resulting exact link (\ref{eqel}) is quite simple and clear.
That is, the growth rate of SR breathers can be found exactly by the absolute difference of group velocities of these two quasi-Akhmediev breathers, i.e., $G_{sr}=\eta_r\Delta V_{gr}$. If $V_{gr1}=-V_{gr2}$, the growth rate of SR breathers, $G_{sr}$, coincides with that of each quasi-Akhmediev breather along the $z$ axis, $2\eta_r|V_{grj}|$ [see $\cosh\Theta_j$ in Eq. (\ref{eqgr})], i.e., $G_{sr}=2\eta_r|V_{grj}|$. This special case covers the previous result obtained in the simplest NLSE \cite{SR2}.


Next we compare and analyze the amplitude amplification of $\delta u$ extracted from the exact solution of SR breathers and
the prediction of linear stability analysis.
The evolution of amplitude maximum $|u|_{max}$ of different SR states as $\Delta V_{gr}\rightarrow 0$ is compared with the
amplitude amplification of linear stability analysis, i.e., $|a+f\exp(G_{sr}\xi)|$ (where $f=\{|u(0,\tau)|-a\}_{max}$).

\begin{figure}[htb]
\centering
\includegraphics[height=32mm,width=84mm]{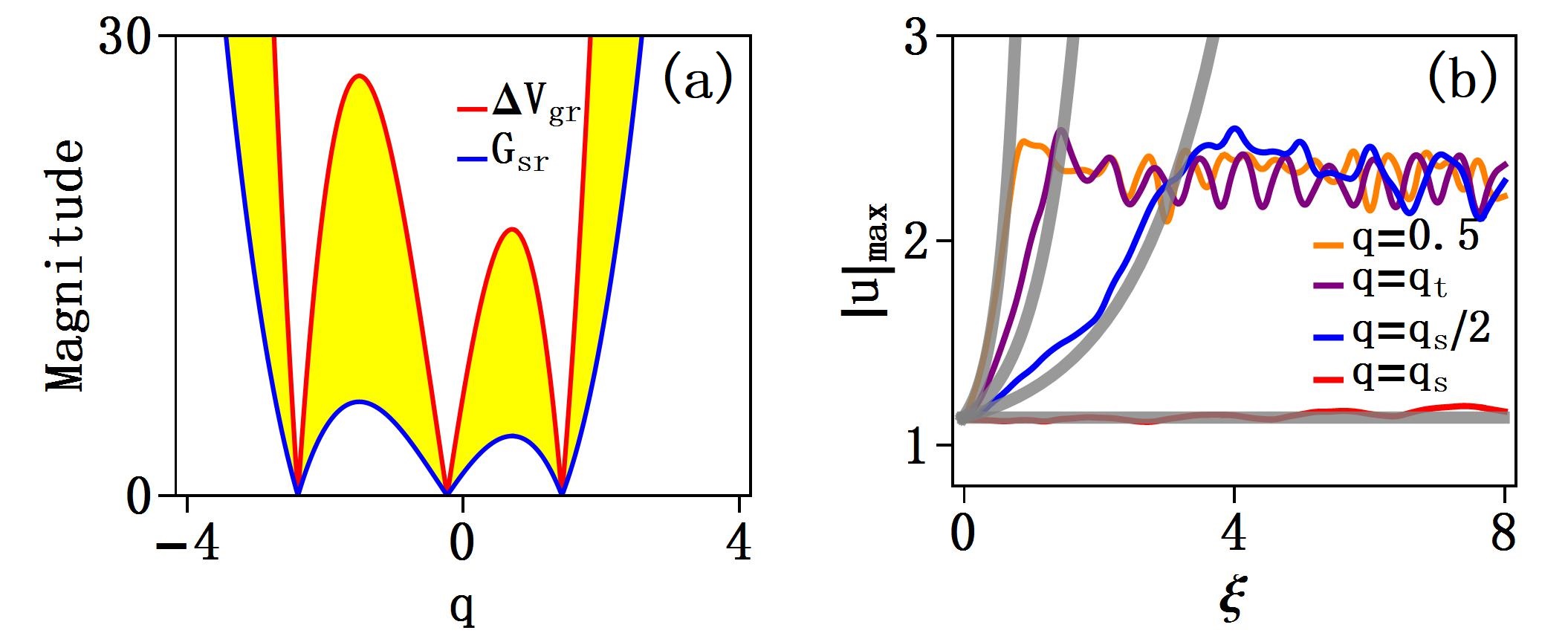}
\caption{(a) Evolution of $\Delta V_{gr}$ Eq. (\ref{eqv1}) and $G_{sr}$ Eq. (\ref{eqg}).
The yellow region represents the factor $\eta_r$, Eq. (\ref{eqel}). (b) Comparison
of amplitude amplification between exact SR solution $|u(\tau,\xi)|_{max}$ and the prediction of linear MI (gray solid lines),
i.e., $|a+f\exp(G_{sr}\xi)|$ (where $f=\{|u(0,\tau)|-a\}_{max}$).
}\label{figli}
\end{figure}

As shown in Fig. \ref{figli}(b),
the evolution of initial amplifying stages of the exact SR breather solution and the linear stability analysis coincide closely.
However, the subsequent stages are totally different. The SR breathers show a nonlinear oscillation after the initial amplification,
while the linear stability analysis becomes completely invalid for the nonlinear stage. Thus the SR breather describes a complete MI scenario
that involves both the linear and nonlinear stages.
These results specifically confirm the MI nature of SR breathers.

\section{extension to Infinite NLSEs}
The exact link (\ref{eqel}) reported above is obtained by the NLSEs with finite orders. An interesting question is that does it hold for the infinite NLSE hierarchy (\ref{eqin}). We address this question in this section.

In principle, one can obtain the general exact SR breather solution by solving the NLSE hierarchy up to infinite order,
step-by-step. However, it is still an open question as to how  a compact solution form for the infinite NLSE (\ref{eqin}) can be obtained. This difficulty comes from the complexity of the higher-order terms. This complexity grows sharply as the order increases \cite{HA1}.
Fortunately, a compact form for the first-order breather solution can be achieved by virtue of the so-called hypergeometric function \cite{HA1}, if one considers the infinite NLSE (\ref{eqin}) with $q=0$.
In this regard, it is still important to find out whether the exact link (\ref{eqel}) is valid for this case.

To this end, we shall first recall the first-order breather solution for Eq. (\ref{eqin}) in Refs. \cite{HA1,HA2} and extract the expression of group velocities $V_{grj}$ with the spectrum parameters parameterized by the Jukowsky transform in Eq. (\ref{eqla}) [see Appendix B]. After a tedious calculation, the absolute difference of group velocity $\Delta V_{gr}$ for the infinite NLSE (\ref{eqin}) is written as
\begin{eqnarray}
\Delta V_{gr}&=&\left|\sum_{n=0}^{\infty}\alpha_{2n+2}\frac{(2n+1)!}{(n!)^2}{}_2F_1\left(1,-n;\frac{3}{2};\sin^2\phi\right)\right|\nonumber\\
&&\times 4A|\sin\phi|,\label{eqinv}
\end{eqnarray}
where ${}_2F_1$ is the hypergeometric function. Clearly, $\Delta V_{gr}$ in Eq. (\ref{eqinv}) has the infinite-order NLSE terms and covers the result in Eq. (\ref{eqv1}) with $q=0$. Note that, however, only even terms contribute to Eq. (\ref{eqinv}). This indicates that if we merely consider the odd terms of Eq. (\ref{eqin}), the SR breather will always exhibit a bound state ($\Delta V_{gr}=0$).

Let us then consider the linear MI growth rate for the infinite NLSE (\ref{eqin}).
Here the plane-wave background is written as
\begin{eqnarray}
u_0=\exp{\left[i\left(\sum_{n=1}^{\infty}\frac{(2n)!}{(n!)^2}\alpha_{2n}\right)\xi\right]}.\label{eqinb}
\end{eqnarray}
Followed by the standard linearization process above, the linear MI growth rate of the plane wave Eq. (\ref{eqinb}) can be obtained as
\begin{eqnarray}
G&=&\left|\sum_{n=0}^{\infty}\alpha_{2n+2}\frac{(2n+1)!}{(n!)^2}{}_2F_1\left(1,-n;\frac{3}{2};\frac{Q^2}{4}\right)\right|\nonumber\\
&&\times \left|Q\sqrt{4-Q^2}\right|,\label{eqingr}
\end{eqnarray}
where $Q$ is the MI perturbed frequency in the range $Q\in(-2,2)$. Note that the linear MI growth rate, Eq. (\ref{eqingr}), can also be obtained by the exact general AB solution for the infinite NLSE (\ref{eqin}), since the initial linear MI is exactly described by the AB solution \cite{AB1,ABb}. Indeed, we find that Eq. (\ref{eqingr}) coincides exactly with the growth factor of the general AB solution in Ref. \cite{HA1}.
If we consider the initial frequency of SR breathers $Q=Q_{sr}=2\sin\phi$, Eq. (\ref{eqingr}) reduces to
\begin{eqnarray}
G_{sr}&=&\left|\sum_{n=0}^{\infty}\alpha_{2n+2}\frac{(2n+1)!}{(n!)^2}{}_2F_1\left(1,-n;\frac{3}{2};\sin^2\phi\right)\right|\nonumber\\
&&\times2|\sin2\phi|.\label{eqingr1}
\end{eqnarray}
Remarkably, in the case of infinite NLSE extensions, the exact relation between SR breathers and MI, Eq. (\ref{eqel}), is still valid, i.e., $G_{sr}=\Delta V_{gr}\eta_r$ [see Eqs. (\ref{eqinv}) and (\ref{eqingr1})].
Namely, the exact link holds for a series of NLSEs up to infinite order. Moreover, one can see from Eqs. (\ref{eqinv}) and (\ref{eqingr1}) that the bound SR breather state ($\Delta V_{gr}=0$) exists when we consider only odd terms of NLSEs, where the MI is suppressed completely ($G_{sr}=0$). This is consistent with the results analyzed in section II. Indeed, this bound state
exhibits the small-amplitude full-suppression SR breather structure which is corresponding to the vanishing growth rate. One particular example can be seen in the simplest complex Korteweg-de Vries model obtained recently \cite{HSR2}.

\section{non-ideal initial excitations}

The initial perturbation $\delta u$ is created by a specially designed collision of multiple quasi-ABs.
However, we find that the form of $\delta u$, Eq. (\ref{eqp}), is representative and illuminating. It implies that
various non-ideal initial perturbations can be used to excite the rich dynamics of SR breathers.
To show this, the ideal perturbation should be replaced by a generalized form:
\begin{equation}
\delta u=-i\{\sum_{j=1}^n L_j(\tau-\tau_j)\cos{\left[Q_j(\tau-\tau_j)+\sigma_j\right]}\}.
\label{eqpn}
\end{equation}
$\delta u$ contains multiple localized perturbations $L_j(\tau-\tau_j)$ with corresponding frequencies of periodic modulation $Q_j$.
Here $L_j$ denotes different types
of localized functions with amplitude $\rho$, width $b$, and time shift $\tau_j$. In general, these parameters allow us to modulate nonideal
initial states to be close to the ideal one readily. The modulated frequency $Q_j$ should approach  the perturbed frequency of SR breathers
(i.e., $Q_j\approx Q_{sr}=2a\sin{\phi}$) in order to excite the SR breather dynamics. One should note that a generalized form of non-ideal initial states
has been used to generate the standard SR breathers in the standard NLSE by solving the Zakharov-Shabat eigenvalue problem \cite{SR4}.
Here, various SR nonlinear evolutions with higher-order effects from non-ideal initial states are verified numerically by the split-step Fourier method. However, the direct numerical simulation for the infinite NLSEs is unrealistic.
Even for the finite-order NLSEs, this should be studied step by step because of the complexity of the higher-order terms.
Thus we confine our attention to the NLSEs up to fourth order, and study both single and multiple SR excitations from Eq. (\ref{eqpn}).
\begin{figure}[htb]
\centering
\includegraphics[height=42mm,width=84mm]{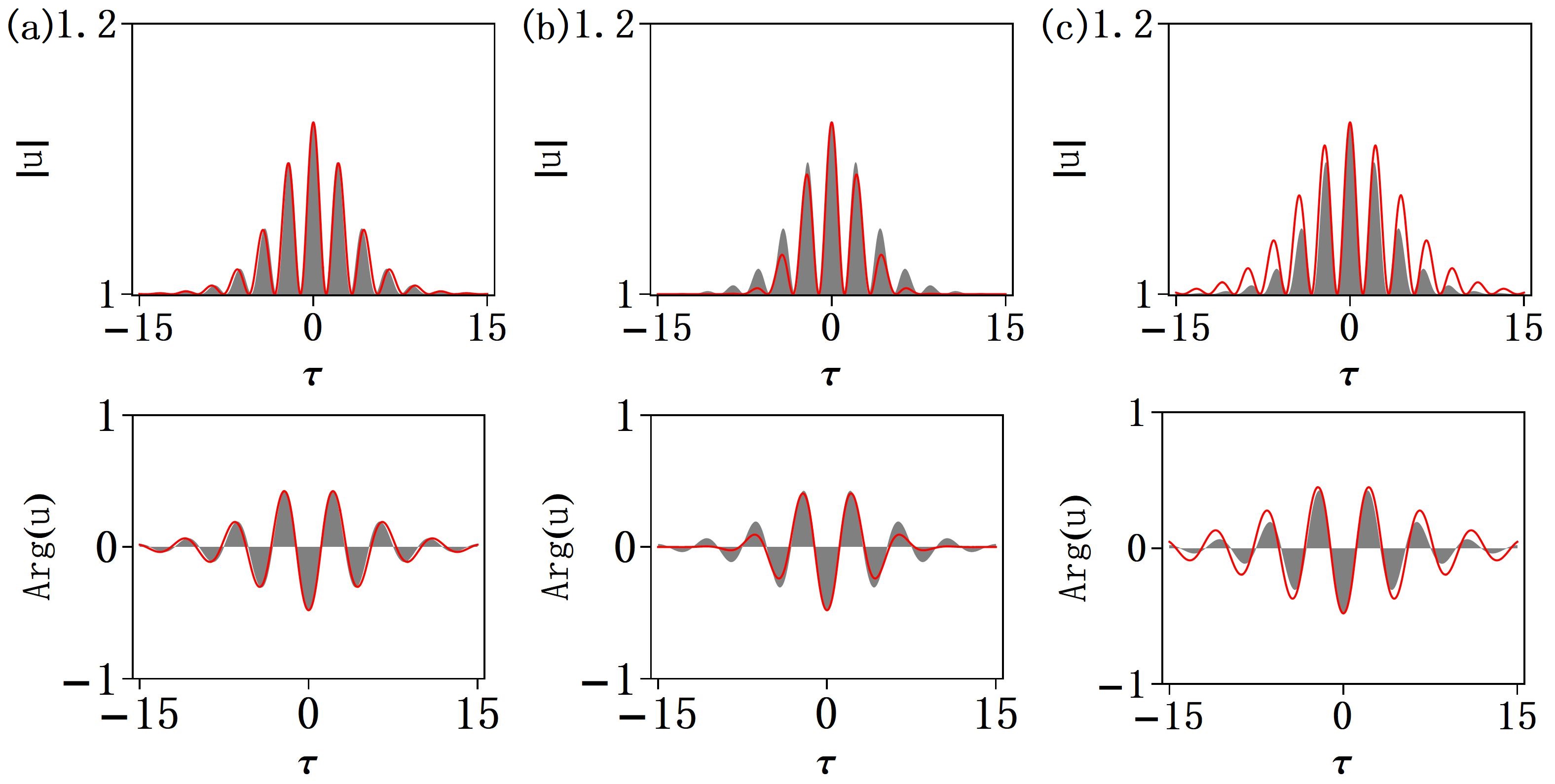}
\caption{(color online)
Amplitude $|u(0,\tau)|$ and phase $\textrm{Arg}[u(0,\tau)]$ profiles (red lines) of nonideal initial pulses,
(a) the sech form $L(\tau)=0.52\textrm{sech}(0.25\tau)$, (b) the Gaussian form $L(\tau)=0.52\exp(-\tau^2/25)$, and (c) the Lorentzian form $L(\tau)=0.52/(1+0.008\tau^2)^2$. The gray regions represent the amplitude and phase of ideal initial state. The setup is the same as in Fig. \ref{figso}, but $\alpha_4=0.01, \alpha_5=0$. Note that the Gaussian and Lorentzian initial states deviate from the ideal initial state.
}\label{figns}
\end{figure}

\begin{figure}[htb]
\centering
\subfigure{\includegraphics[height=24mm,width=27mm]{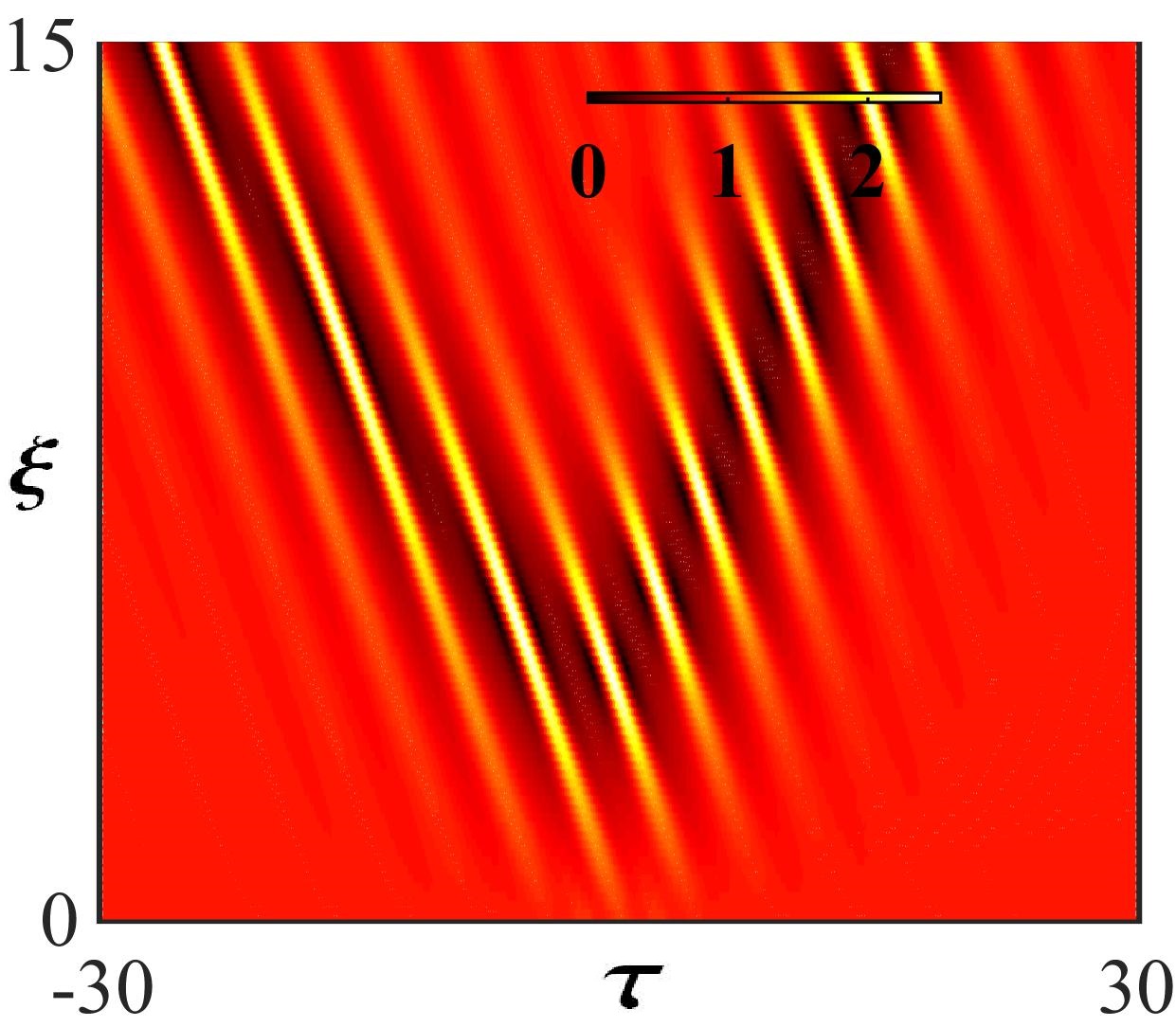}}
\subfigure{\includegraphics[height=24mm,width=27mm]{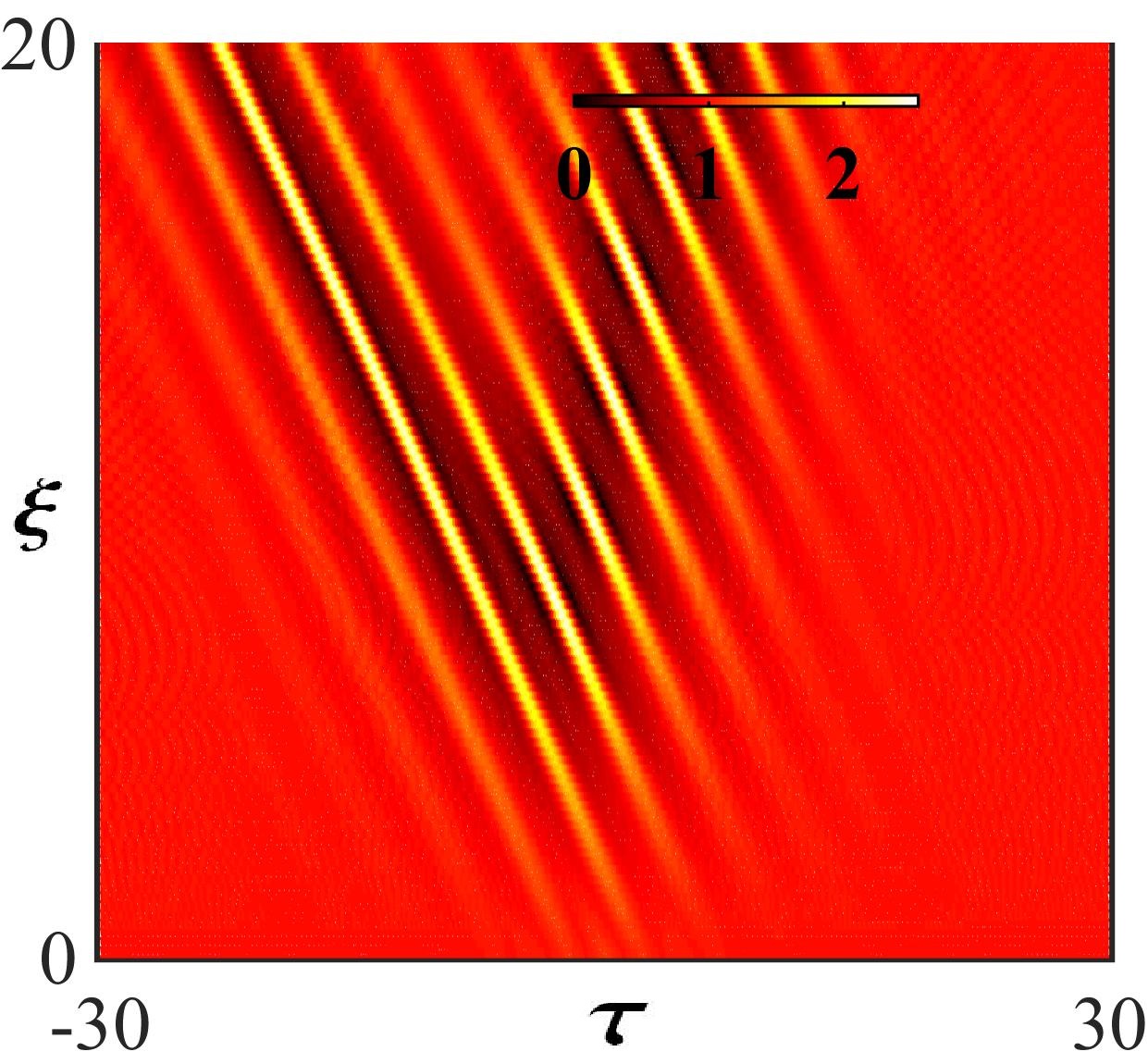}}
\subfigure{\includegraphics[height=24mm,width=27mm]{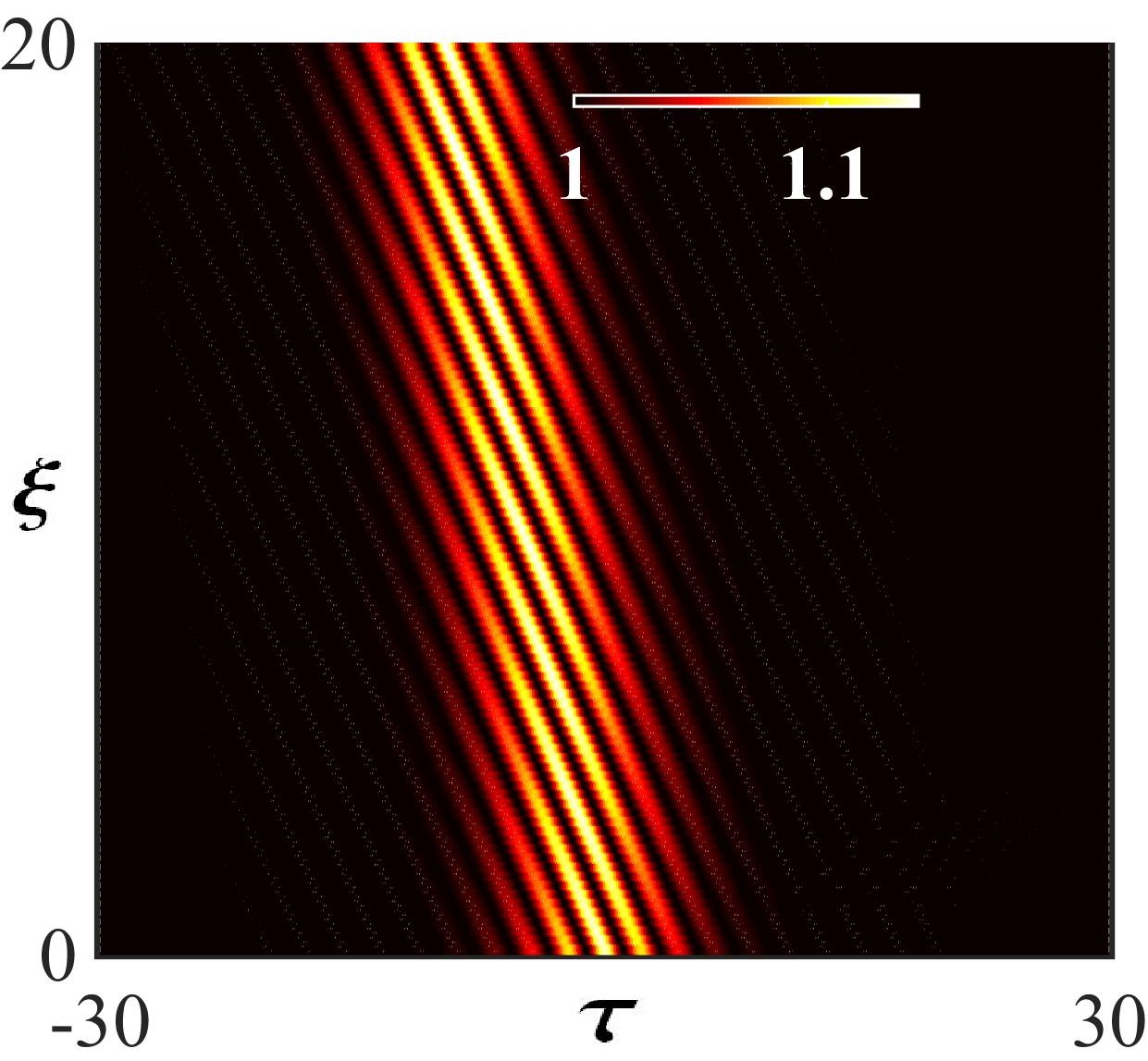}}
\subfigure{\includegraphics[height=24mm,width=27mm]{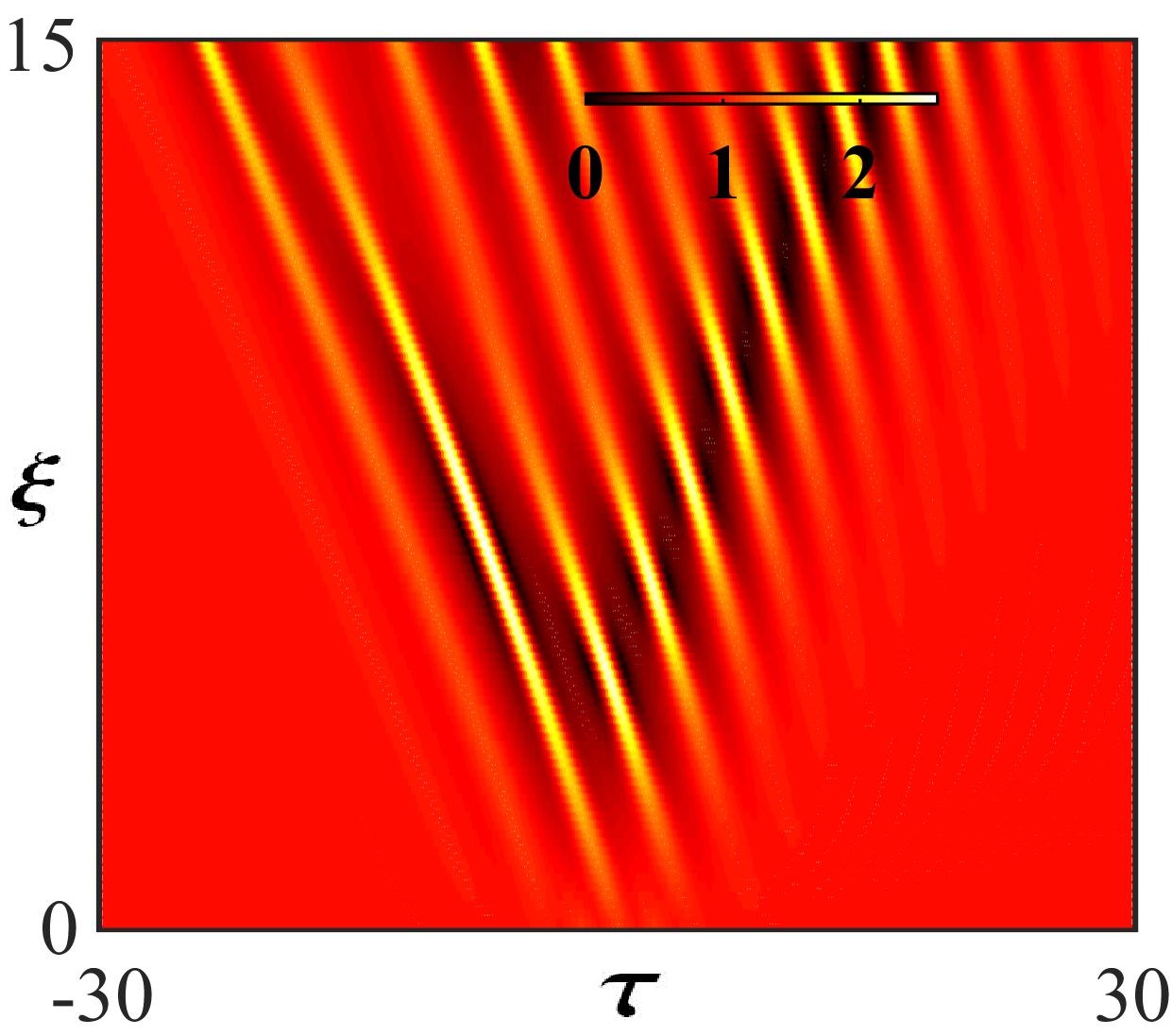}}
\subfigure{\includegraphics[height=24mm,width=27mm]{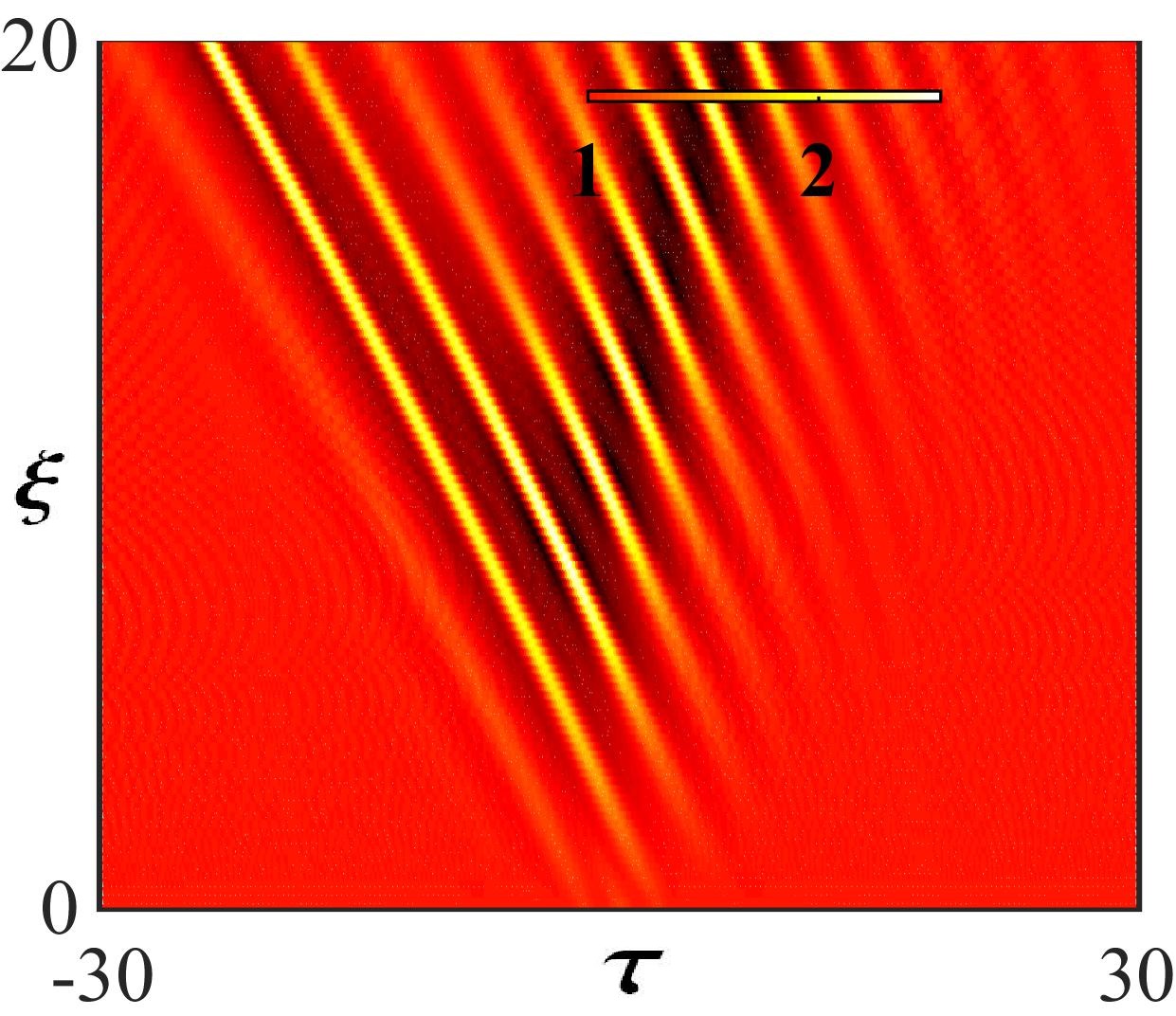}}
\subfigure{\includegraphics[height=24mm,width=27mm]{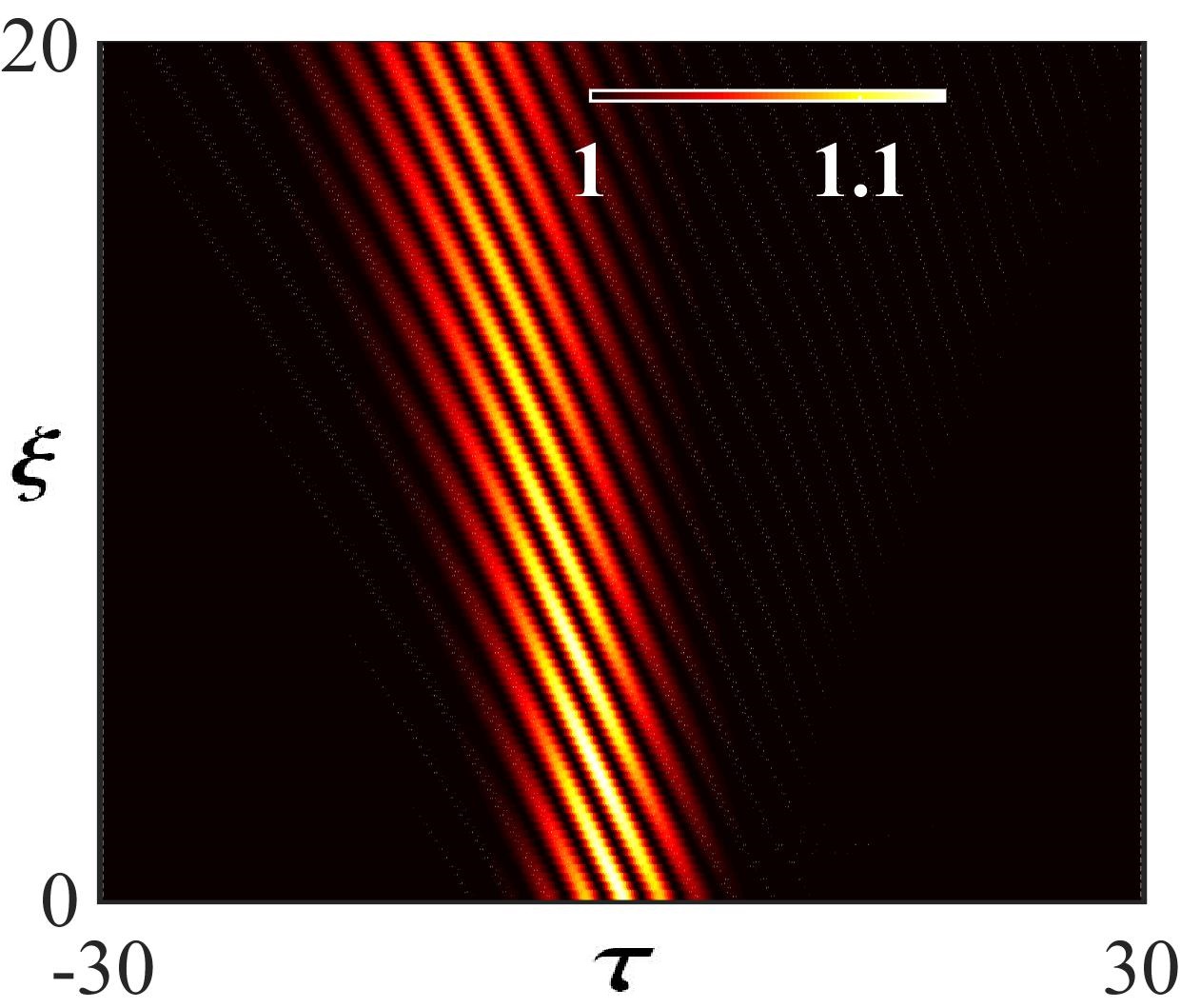}}
\subfigure{\includegraphics[height=24mm,width=27mm]{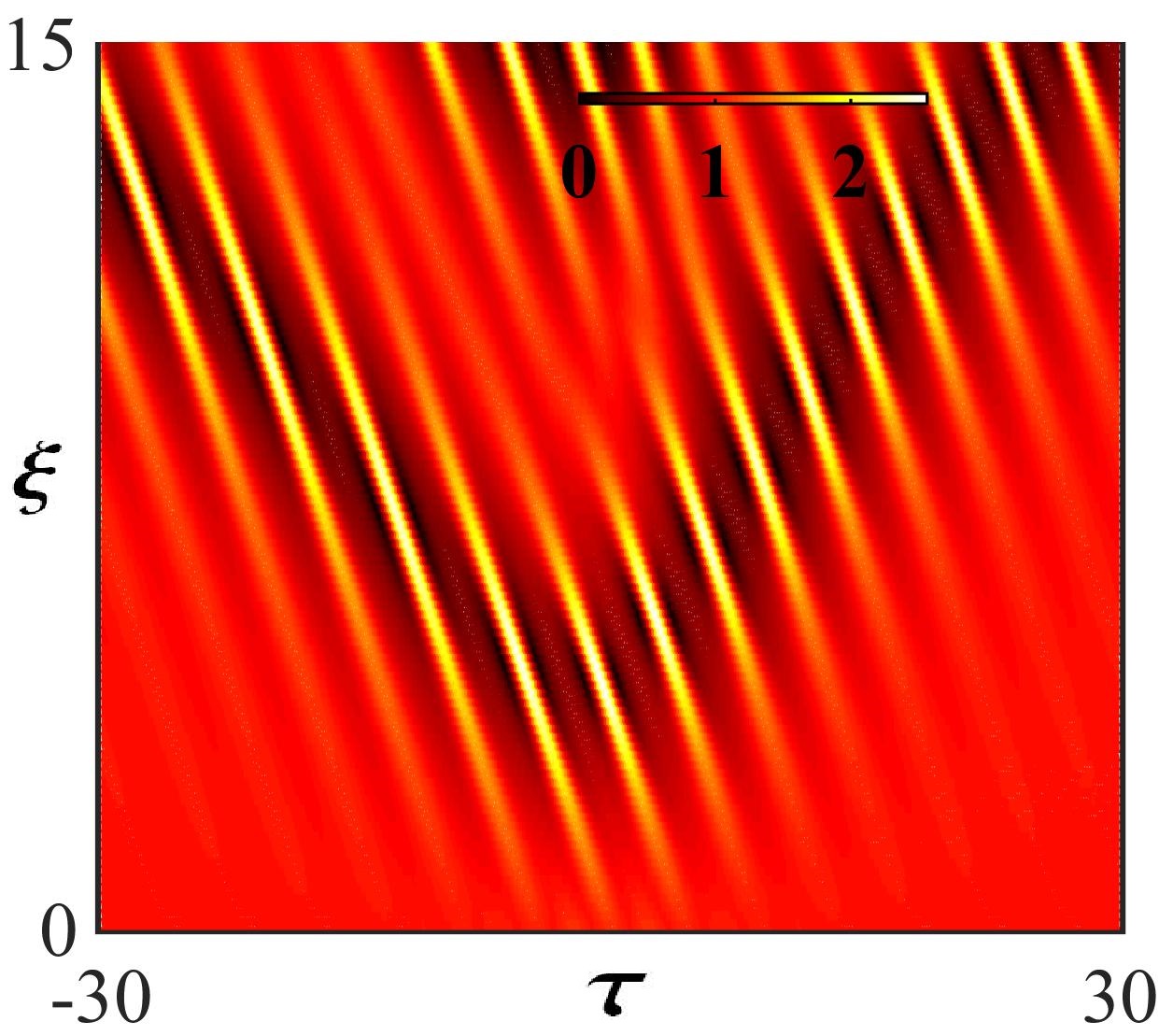}}
\subfigure{\includegraphics[height=24mm,width=27mm]{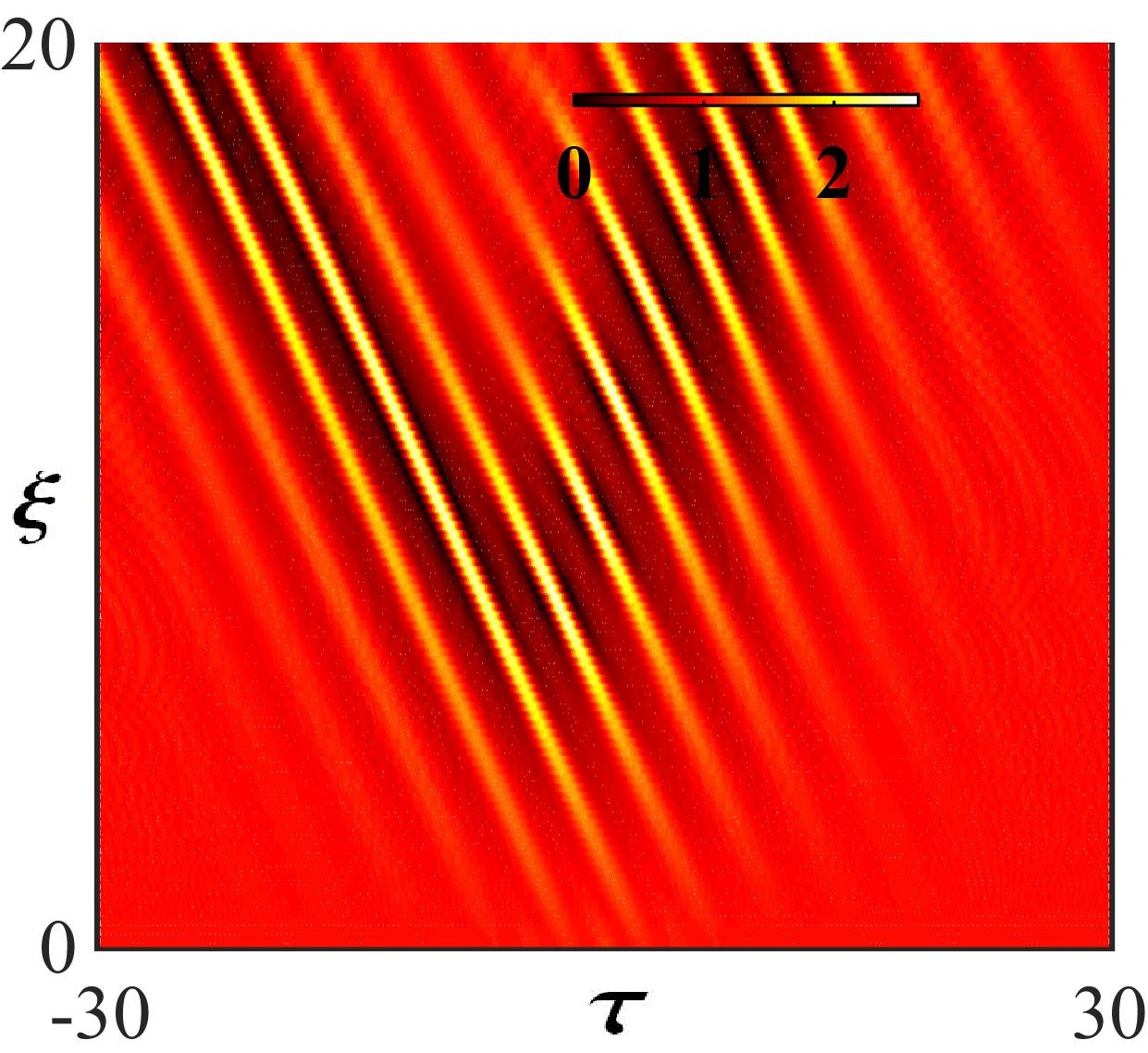}}
\subfigure{\includegraphics[height=24mm,width=27mm]{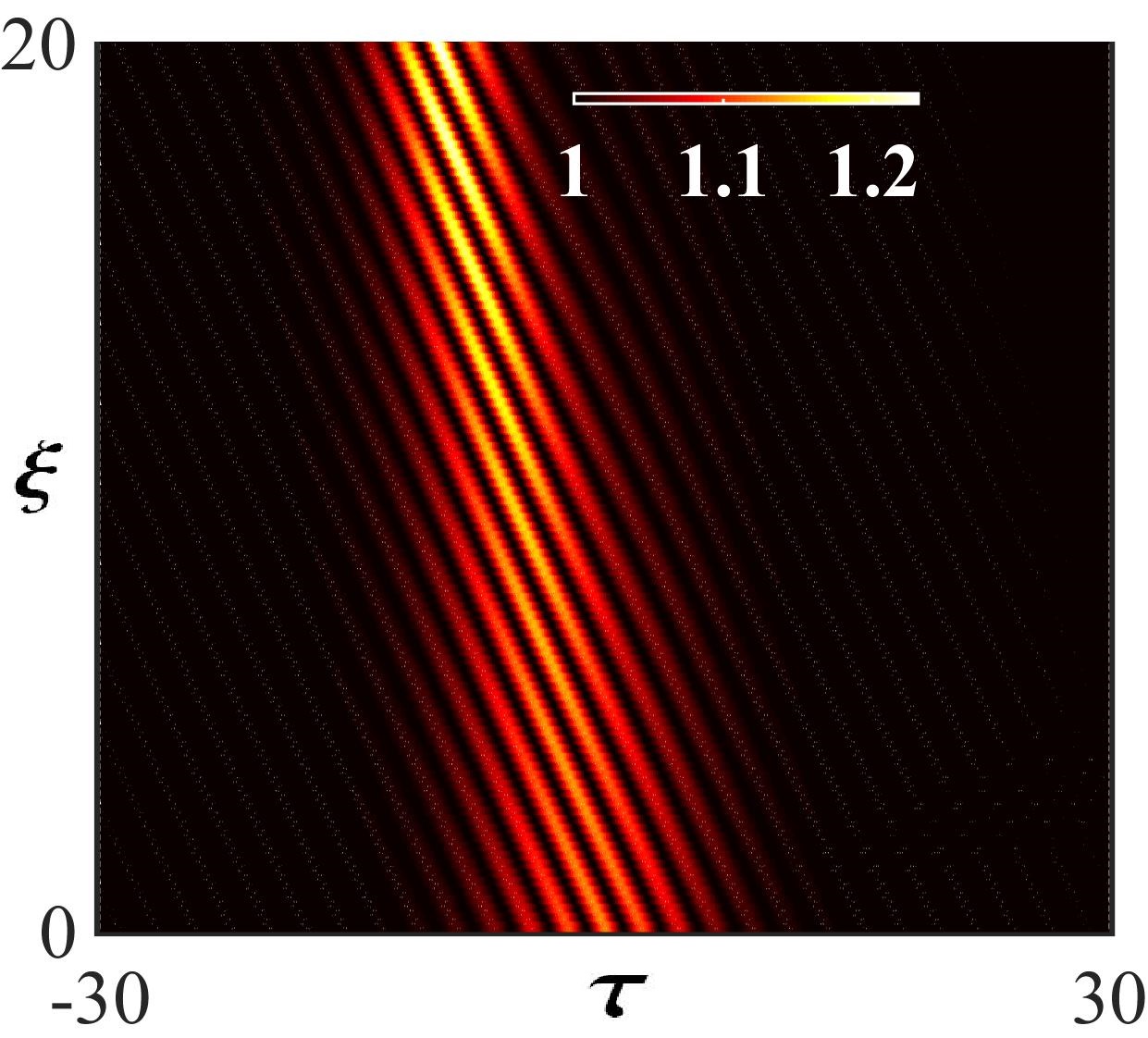}}
\caption{Numerical simulations $|u(\xi,\tau)|$ of different SR states [standard SR breather (left row); half-transition SR breather (middle column); full-suppression bound SR breather (right column)] from non-ideal initial states in Fig. \ref{figns}: the `sech'- type (top row); the Gaussian type (middle row); the Lorentzian type (bottom row).  Accurate
 reappearance of different SR breather dynamics can be observed numerically, although the Gaussian and Lorentzian initial states deviate from the ideal initial state.
}\label{figns1}
\end{figure}
\begin{figure}[htb]
\centering
\subfigure{\includegraphics[height=24mm,width=27mm]{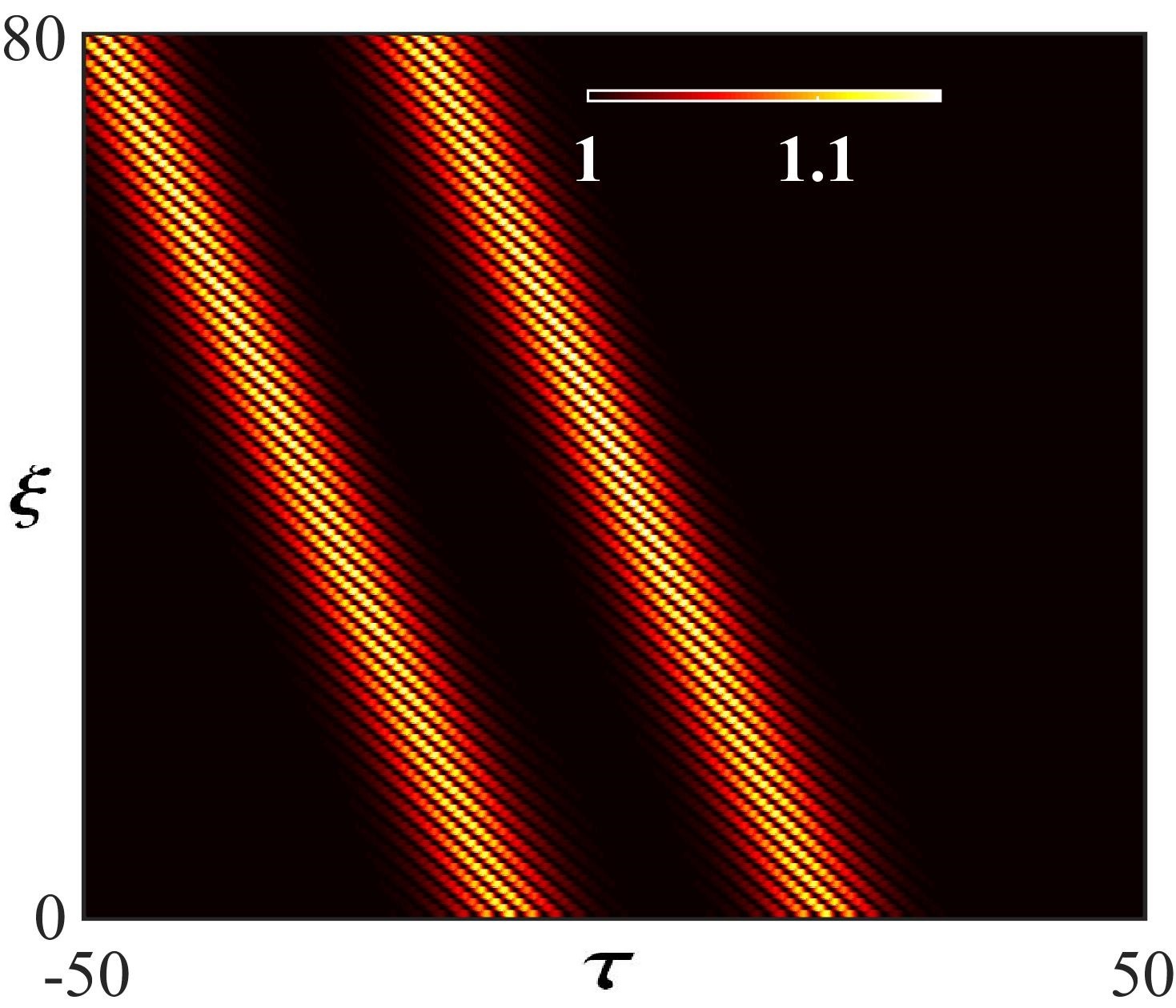}}
\subfigure{\includegraphics[height=24mm,width=27mm]{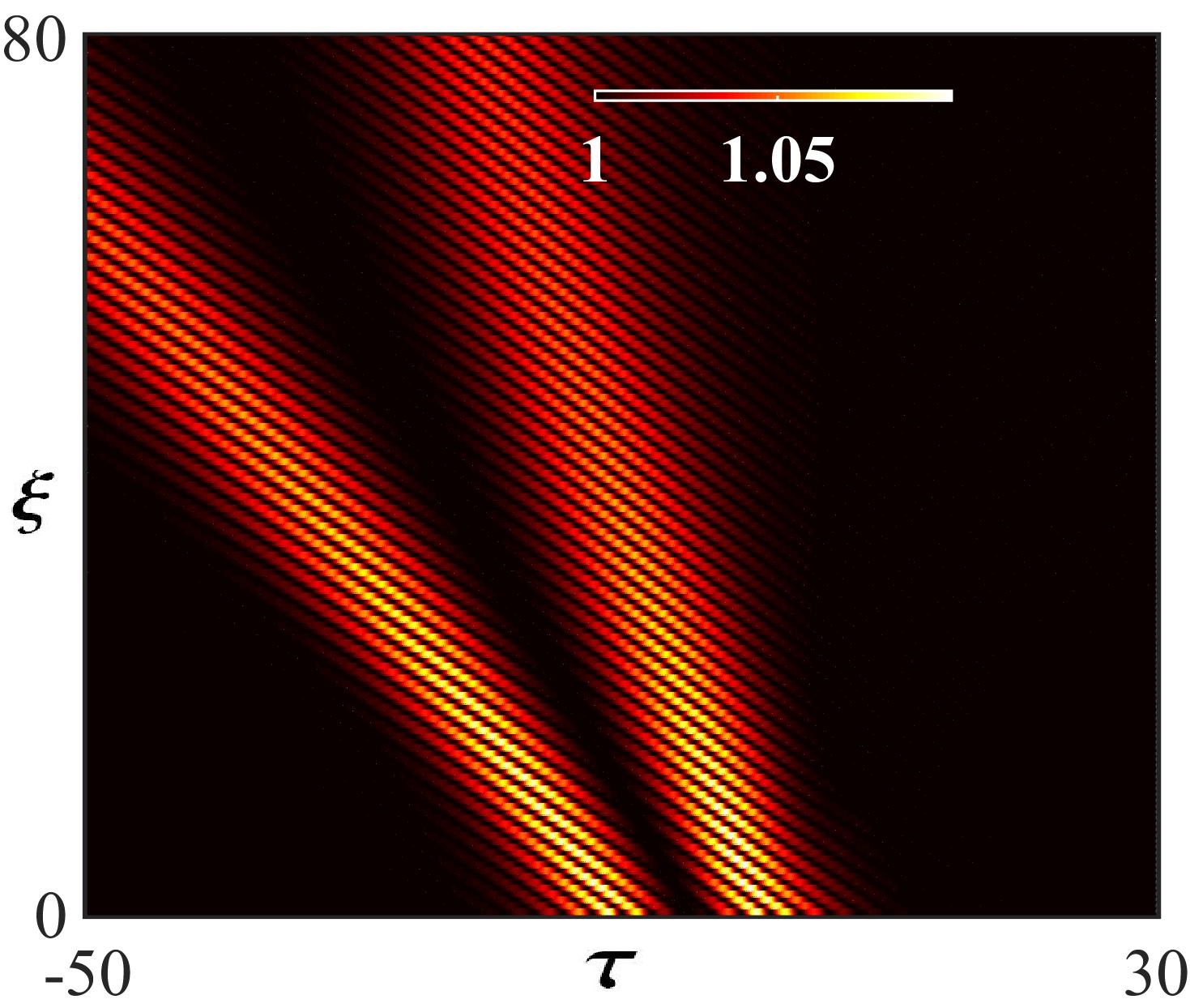}}
\subfigure{\includegraphics[height=24mm,width=27mm]{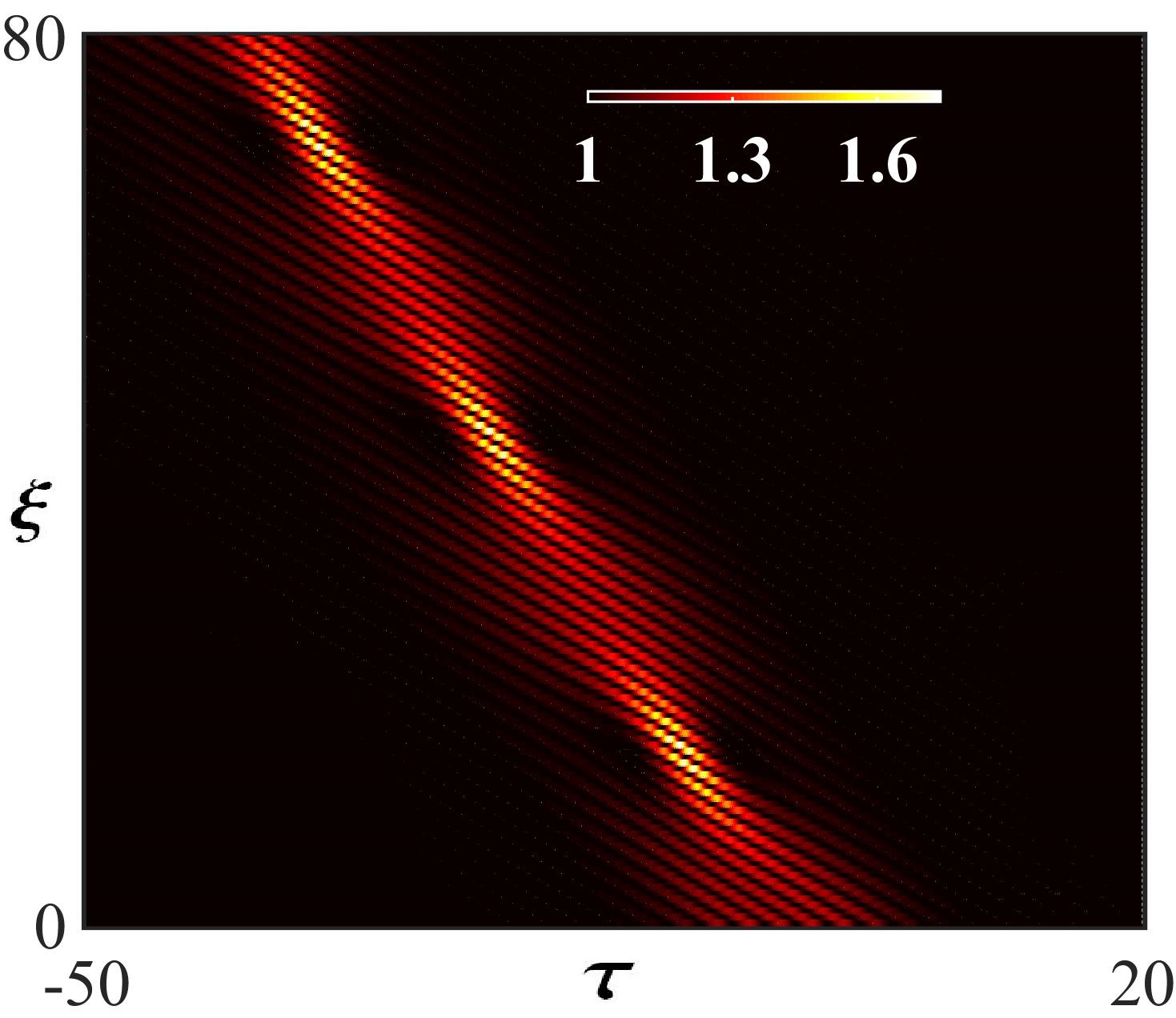}}
\caption{Numerical simulations $|u(\xi,\tau)|$ of double excitation of SR bound states from initial perturbation $\delta u=-i \{L_1(\tau-\tau_1)\cos{\left[Q_1(\tau-\tau_1)+\sigma_1\right]}+L_2(\tau)\cos{\left[Q_2(\tau)\right]}\}$ with $\sigma_1=\pi/2$ as $\tau_1$ decreases (from left to right $\tau_1=30,10,7$). Here $L_{1}=0.52\textrm{sech}[0.25(\tau-\tau_1)]$, $L_{2}=0.52\textrm{sech}(0.25\tau)$.
}\label{figns2}
\end{figure}

\subsection{single initial excitation}
First, we choose various simple single pulse forms that are quite convenient for operation in practice, including the `sech' type $L(\tau)=\rho\,\textrm{sech}(b\tau)$, Gaussian-type $L(\tau)=\rho \exp(-\tau^2/b)$, and Lorentzian-type $L(\tau)=\rho/(1+b\tau^2)^2$. However, we choose non-ideal pulses with quite different profiles [see Fig. \ref{figns}]
in order to show the universality of SR breather dynamics.
As shown in Fig. \ref{figns}, three different types of pulses are chosen as initial states, i.e, the sech, Gaussian, and Lorentzian forms.
Only the `sech' form exhibits   consistency  with the exact one. The Gaussian and Lorentzian initial states deviate from the exact initial state; the Gaussian-type is narrower and has fewer  peaks while the Lorentzian type is wider and has more peaks compared with the the exact state.
Fig. \ref{figns1} shows the nonlinear dynamics from different initial states in Fig. \ref{figns}.

Importantly, all these non-ideal initial states can create  striking dynamics in non-trivial SR breathers.
In particular, the SR breather generated from the `sech' pulse (the non-ideal initial state which is consistent with exact initial state) is almost the same as the spatial-temporal distribution of the exact solution [see Fig. \ref{figns1}(top row)].
The others (Gaussian and Lorentzian states beyond exact initial state) can still create robust SR breather  dynamics as $\xi$ increases. However, the resulting nonlinear states exhibit slight differences, compared with the exact SR nonlinear state.

Specifically, the standard SR breathers that have evolved from the small-size Gaussian and large-size Lorentzian perturbations show more complex nonlinear states with smaller and larger transverse distributions, respectively. In particular,
a nonlinear recurrence of quasi-ABs at a larger propagation distance $\xi=15$ is observed. This phenomenon could be categorized as a higher-order MI \cite{MIH1}, which comes from the initial state beyond the critical value of exact initial state.

For the half-transition case, however, we note that the recurrence phenomenon vanishes at the nonlinear stage. The half-transition SR states show greater robustness than the standard SR breathers. Also, the small-size Gaussian state leads to SR breathers with a smaller transverse distribution, while the large-size Lorentzian state induces SR breathers with a larger transverse distribution. Note that in the half-transition SR case, the small- and large-size initial states result in multi-peak solitons with fewer  and more peaks, respectively.

On the other hand, for the full-suppression bound state case, both Gaussian and Lorentzian initial states with small amplitudes propagate stably for the first ten propagation units. After that, the small-size Gaussian pulse becomes broadened and weakens while the large-size Lorentzian pulse is narrowed and amplified slightly. Nevertheless, we note that both of them cannot be amplified greatly over a large propagation distance, since all these states exist in the modulation stability regime with vanishing growth rate; the small-size Gaussian pulse is broadened and the large-size Lorentzian pulse shows a long-distance periodic structure with small amplitude.

\subsection{multiple initial excitations}
Next, we consider how  multiple excitation of SR waves can be realized, when one considers an initial state with multiple localized perturbations.
This has been used to generate higher-order rogue waves, analytically and numerically \cite{SR4}.
We note that the dynamics of the multiple excitation of SR waves in Eq. (\ref{eqin}) is  richer, since SR waves exhibit structural diversity with higher-order effects.
Here we only show the double excitation of full-suppression SR bound states with $\delta u=-i \{L_1(\tau-\tau_1)\cos{\left[Q_1(\tau-\tau_1)+\sigma_1\right]}+L_2(\tau)\cos{\left[Q_2(\tau)\right]}\}$. Thus $\tau_1$ and $\sigma_1$ are the relative shift and phase. If $\tau_1$ is large, a  parallel double excitation of SR bound states will be observed with arbitrary $\sigma_1$ [see Fig. \ref{figns2} (left)].
However, as $\tau_1$ decreases, the nonlinear dynamics becomes complex, and depends  on the values of $\tau_1$ and $\sigma_1$. As shown in Fig. \ref{figns2},  repelling and periodic beating structures emerge as $\tau_1$ decreases, when we fix $\sigma_1=\pi/2$.
One should keep in mind that the multiple excitation of SR bound states appear in the regime with a vanishing growth rate. Thus the corresponding dynamics is similar to the one of classical bright solitons \cite{Chen}, although the SR breather bound states propagate on a nonzero background.
This could be useful  for  stable excitations of SR breather bound states in mode-locked laser, where various soliton `molecules' \cite{Liu} and the internal dynamics of femtosecond soliton `molecules' \cite{BS} have  recently  been demonstrated.
It is expected that these results would  confirm the robustness and universality of SR breather dynamics with higher-order effects and would  greatly broaden the applicability of SR breathers in  related nonlinear physics.

\section{Conclusions and discussions}
In summary, we have addressed a major problem for the ubiquitous MI{---}the general exact link between Zakharov-Gelash SR breathers and the
linear MI. It confirms that the absolute difference of group velocities of SR breathers coincides with the MI growth rate.
This crucial link holds for a series of NLSEs up to infinite order.
The universality of different SR breathers is revealed numerically by showing the robustness of SR breather evolution from various non-ideal initial excitations. These results shed light on the nature of MI described by the SR breather theory.
Similar studies can be extended to various multi-component systems
in which vector nonlinear waves usually exhibit much richer dynamics \cite{v1,v2,v3,v4,v5,vsr}. It is expected that vector SR breathers will describe
a broad range of MI scenarios in complex nonlinear physics. Recently, SR breathers in a special multi-component system, i.e., the multiple self-induced transparency system have been demonstrated \cite{vsr}.

The MI evolution from non-ideal initial states is still a challenging work in nonlinear science. Our numerical result can be regarded as the first step of this general problem. The next step should include the analysis of the eigenvalue spectrum of the non-ideal initial state \cite{SR4} and a linear stability analysis of the perturbation caused by a non-ideal initial condition \cite{AC}.

The crucial exact link (\ref{eqel}) obtained in this paper can stimulate the interest for the SR waves in many non-integrable nonlinear systems. Without exact SR breather solutions in this case, however, a linear stability analysis for a plane wave background can be achieved easily. Thus, it is expected that the MI dynamics described by the SR breather theory could be well controlled based on the exact link (\ref{eqel}). One interesting work is to study the ``quartic SR breather'' (i.e., the fourth-order dispersion is the dominant dispersion effect) in a specially-designed system of photonic crystal waveguides, where the ``pure-quartic soliton'' has been demonstrated numerically and experimentally \cite{He}. Thanks to
this general exact link (\ref{eqel}), novel ``quartic SR breathers'' (e.g., the bound SR wave state predicted above) could be observed by the modulation of the linear MI growth rate.

Recently, special attentions have been paid on on various scenarios of nonlinear MI and nonlinear dynamics from different initial states. These include  Fermi-Pasta-Ulam recurrence \cite{FPU}, integrable turbulence with random perturbations \cite{IT}, heteroclinic modes \cite{Conforti}, longtime asymptotic states associated with a continuous spectrum \cite{EI,cs,Biondini,Biondini1}, and doubly-periodic structures \cite{JM}.
The links and differences between them remain widely unexplored.
However, their different dynamics manifestations will enrich our understanding of MI and nonlinear dynamics.
Indeed, even for a simple localized (i.e., Gaussian) perturbation with a purely imaginary form in the standard NLSE, different modulated parameters ($\rho,b$)
of the initial perturbation can led to distinct nonlinear stages. Namely, a SR breather dynamics can be readily created when
the initial Gaussian state is comparable with that of exact SR breathers (see Fig. \ref{figns1} and also Ref. \cite{SR4}), while a
formation of `sech'-shaped soliton trains
associated with a continuous spectrum will be generated when one uses a strong Gaussian initial state ($\rho=b=1$) in Ref. \cite{Biondini1}.
Note also that the similar broadening of the solution with a constant drift velocity of the edge of the perturbation has been reported in \cite{EI}.
Is there is a general and unified mechanism governing local perturbations of plane waves in NLS equation (independently of the Zakharov-Shabat spectrum)?
The underlying mechanism for this difference is an intriguing question which will be studied in the future.

\section*{ACKNOWLEDGEMENTS}
We are grateful to Prof. Nail Akhmediev and Dr. Adrian Ankiewicz for their many valuable discussions on MI and the extension of results into the infinite NLSEs.
We thank Prof. Spyridon Kamvissis for his valuable comments on MI and longtime asymptotic states of
continuous spectrum. 
This work has been supported by the National Natural Science Foundation of China (NSFC)
(Grant Nos. 11705145, 11475135, 11547302, 11434013, 11775178, and 11425522), the Scientific Research Program Funded by Shaanxi
Provincial Education Department (Grant No.17JK0767), Natural Science Basic Research Plan in Shaanxi Province of China (Grant No. 2018JQ1003), and the Major Basic Research Program of Natural Science of
Shaanxi Province (Grant Nos. 2017KCT-12, 2017ZDJC-32).

\section*{APPENDIX A}
We consider the SR breathers formed by a pair of breathers with $R_1=R_2=R=1+\varepsilon$, $\phi_1=-\phi_2=\phi$.
Then the complex spectral parameters are
\begin{equation}
\lambda_j=i\frac{1}{2} a \left(R+1/R\right)\cos \phi\mp\frac{1}{2} \left[a \left(R-1/R\right) \sin \phi+q\right].\label{eql1}
\end{equation}
The corresponding analytical SR breather solution for Eq. (\ref{eqin}) with first five terms can be obtained via the iteration of Darboux transformation
which is of the form:
\begin{eqnarray}
u=u_0\left(1-4\rho \varrho_1\frac{(i\varrho_1-\rho)\Xi_1
+(i\varrho_1+\rho)\Xi_2}
{a(\rho^2\Xi_3+\varrho_1^2\Xi_4)}\right),\label{equsr}
\end{eqnarray}
where $\varrho_1=\frac{a}{2}\left(R-1/R\right)\sin{\phi},~\rho=\frac{a}{2}\left(R+1/R\right)\cos{\phi}$,
\begin{eqnarray}
\Xi_1&=&\varphi_{21}\phi_{11}+\varphi_{22}\phi_{21},~\Xi_2=\varphi_{11}\phi_{21}+\varphi_{21}\phi_{22},\nonumber\\
\Xi_3&=&\varphi_{11}\phi_{22}-\varphi_{21}\phi_{12}-\varphi_{12}\phi_{21}+\varphi_{22}\phi_{11},\nonumber\\
\Xi_4&=&(\varphi_{11}+\varphi_{22})(\phi_{11}+\phi_{22}).\nonumber
\end{eqnarray}
Here $\phi_{jj}$, $\varphi_{jj}$, $\phi_{j3-j}$, and $\varphi_{j3-j}$ are linear combinations of trigonometric
and hyperbolic functions
\begin{eqnarray}
\phi_{jj}&=&\cosh(\Theta_2\mp i\psi)-\cos(\Phi_2\mp\phi),\nonumber\\
\varphi_{jj}&=&\cosh(\Theta_1\mp i\psi)-\cos(\Phi_1\pm\phi),\nonumber\\
\phi_{j3-j}&=&\pm i\cosh(\Theta_2\mp i\phi)\mp i\cos(\Phi_2\mp \psi),\nonumber\\
\varphi_{j3-j}&=&\pm i\cosh(\Theta_1\pm i\phi)\mp i\cos(\Phi_1\mp\psi),\nonumber
\end{eqnarray}
where $\psi=\arctan[(i-iR^2)/(1+R^2)]$.
$\Theta_j$ and $\Phi_j$ contain the group and phase velocities
(i.e., $V_{grj}$ and $V_{phj}$) of the localized wave structures as well as the important free phase parameters (i.e., $\mu_j$ and $\theta_j$),
which are given by
\begin{equation}
\Theta_j=2\eta_{r}(\tau-V_{grj}\xi)+\mu_j,~\Phi_j=2\eta_{ij}(\tau-V_{phj}\xi)-\theta_j,\label{eqgr}
\end{equation}
where $\eta_{i1}=-\eta_{i2}=\eta_{i}$,
$\eta_r=\frac{a}{2}\left(R-1/R\right)\cos{\phi},~\eta_i=\frac{a}{2}\left(R+1/R\right)\sin{\phi}$, and
$V_{grj}=v_{1j}+v_{2j}\eta_{ij}/\eta_{r}, ~~V_{phj}=v_{1j}-v_{2j}\eta_{r}/\eta_{ij}$.
Here $v_{1j}$ and $v_{2j}$ are determined by the coefficients of NLSEs
\begin{eqnarray}
v_{1j}&=&2\alpha_2(q+\varrho_j)-\alpha_3 \chi_{1j}-\alpha_4 \chi_{2j}-\alpha_5 \chi_{3j},\\
v_{2j}&=&\rho(2\alpha_2+\alpha_3 \kappa_{1j}+\alpha_4 \kappa_{2j}+\alpha_5 \kappa_{3j}),
\end{eqnarray}
where $\varrho_2=-\varrho_1$, and
\begin{eqnarray}
\chi_{1j}&=&2a^2-3q^2+4\rho^2-6q\varrho_j-4\varrho_j^2,\nonumber\\
\chi_{2j}&=&4q^3-8a^2q-4a^2 \varrho_j+12q^2 \varrho_j-24\rho^2 \varrho_j+8\varrho_j^3, \nonumber\\
\chi_{3j}&=&6 a^4-20 a^2 q^2+8 a^2 \rho ^2-20 a^2 q \varrho_j-8 a^2 \varrho_j^2+5 q^4\nonumber\\
&&+16 \rho ^4+20 q^3 \varrho_j-40 q^2 \rho ^2+40 q^2 \varrho_j^2-96 \rho ^2 \varrho_j^2\nonumber\\
&&-120 q \rho ^2 \varrho_j+40 q \varrho_j^3+16 \varrho_j^4,\nonumber\\
\kappa_{1j}&=&6 q+8 \varrho_j, \nonumber\\
\kappa_{2j}&=&4 a^2-12 q^2+8 \rho ^2-32 q \varrho_j-24 \varrho_j^2, \nonumber\\
\kappa_{3j}&=&20 a^2 q+16 a^2 \varrho_j-20 q^3-80 q^2 \varrho_j+64 \rho ^2 \varrho_j\nonumber\\
&&+40 \rho ^2 q-120 q \varrho_j^2-64 \varrho_j^3.\nonumber
\end{eqnarray}
\section*{APPENDIX B}
The general expression of group velocity of breather solutions for the infinite NLSE (\ref{eqin}) with the spectrum parameters parameterized by the Jukowsky transform Eq. (\ref{eqla}) is obtained as
\begin{eqnarray}
V_{grj}=v_{1j}+v_{2j}\eta_{ij}/\eta_{r},
\end{eqnarray}
where $v_{1j}=\textrm{Im}\{\vartheta_j\}$, $v_{2j}=\textrm{Re}\{\vartheta_j\}$, respectively, with 
\begin{eqnarray}
\vartheta_j&=&2i\lambda_j\sum_{n=0}^{\infty}\alpha_{2n+2}\frac{(2n+1)!}{(n!)^2}{}_2F_1\left(1,-n;\frac{3}{2};1+\lambda_j^2\right)\nonumber\\
&&+i\sum_{n=1}^{\infty}\alpha_{2n+1}\frac{(2n+1)!}{(n!)^2}{}_2F_1\left(1,-n;\frac{3}{2};1+\lambda_j^2\right).\nonumber
\end{eqnarray}
Here ${}_2F_1$ is the hypergeometric function, $\lambda_j=i\frac{1}{2} \left(R+1/R\right)\cos \phi\mp\frac{1}{2}\left(R-1/R\right) \sin \phi$ with $a=1,~q=0$, $\eta_{r}=\frac{1}{2}\left(R-1/R\right)\cos{\phi}$, and $\eta_{ij}=\mp\frac{1}{2}\left(R+1/R\right)\sin{\phi}$.
Considering the SR breather condition $R=1+\varepsilon$, $\varepsilon\ll1$, we obtain $\lambda_j=i\cos \phi\mp\frac{1}{2}\left(R-1/R\right) \sin \phi$, $\eta_{ij}=\mp\sin{\phi}$, and $1+\lambda_j^2=\sin^2\phi\mp2i\eta_{rj}\sin\phi$.

\end{document}